\documentclass[aps,twocolumn,article,showpacs,showkeys]{revtex4}

\usepackage{amsmath,amssymb,mathrsfs}
\usepackage{latexsym}
\usepackage{graphicx,psfrag} 

\newcommand{\bs}[1]{\boldsymbol{#1}}
\newcommand{\bigc}[2]{\bigl[#1,#2\bigr]}

\newcommand{\comm}[2]{\left[#1,#2\right]}

\newcommand{\bond}{\left\langle i, j \right\rangle}
\newcommand{\vac}{\left|\,0\,\right\rangle}
\newcommand{\ket}[1]{\left|#1\right\rangle}
\newcommand{\bra}[1]{\left\langle#1\right|}
\newcommand{\braket}[1]{\bigl\langle#1\bigr\rangle}
\newcommand{\up}{\uparrow}
\newcommand{\dw}{\downarrow}
\def\bd{\begin{displaymath}}
\def\ed{\end{displaymath}}
\def\be{\begin{equation}}
\def\ee{\end{equation}}
\def\bea{\begin{eqnarray}}
\def\eea{\end{eqnarray}}
\def\bi{\begin{itemize}}
\def\ei{\end{itemize}}
\def\bn{\begin{enumerate}}
\def\en{\end{enumerate}}

\def\ie{{\it i.e.},\ }
\def\eg{{\it e.g.}\ }

\begin{document}
\title{Is electromagnetic gauge invariance spontaneously violated 
in superconductors?}

\author{Martin Greiter} 
\email{greiter@tkm.uni-karlsruhe.de}
\affiliation{Institut f\"ur Theorie der Kond.\ Materie,
  Universit\"at Karlsruhe, Postfach 6980, D-76128 Karlsruhe}

\begin{abstract}
  We aim to give a pedagogical introduction to those elementary
  aspects of superconductivity which are not treated in the classic
  textbooks.  In particular, we emphasize that global U(1) phase
  rotation symmetry, and not gauge symmetry, is spontaneously
  violated, and show that the BCS wave function is, contrary to claims
  in the literature, fully gauge invariant.  We discuss the nature of
  the order parameter, the physical origin of the many degenerate
  states, and the relation between formulations of superconductivity
  with fixed particle numbers {\it vs.}\ well defined phases.  We
  motivate and to some extend derive the effective field theory at low
  temperatures, explore symmetries and conservation laws, and justify the
  classical nature of the theory.  Most importantly, we show that the
  entire phenomenology of superconductivity essentially follows from
  the single assumption of a charged order parameter field.  This
  phenomenology includes Anderson's characteristic equations of
  superfluidity, electric and magnetic screening, the Bernoulli Hall
  effect, the balance of the Lorentz force, as well as the quantum
  effects, in which Planck's constant manifests itself through the
  compactness of the U(1) phase field.  The latter effects include
  flux quantization, phase slippage, and the Josephson effect.
% \\  PACS numbers: 74.20.-z, 11.15.-q, 11.15.Ex.%~\cite{pacs}
\end{abstract}



\maketitle
%\tableofcontents

\section*{CONTENTS}
\noindent{\small
\begin{tabular}{r@{\hspace{2mm}}p{75mm}r}
I. &Introduction &1\\[3pt]
II. &Gauge invariance &2\\[3pt]
III. &Order parameter considerations &4\\[3pt]
IV. &Effective field theory &6\\[3pt]
V. &Phenomenology and the Higgs mechanism &11\\[3pt]
VI. &Quantum effects &15\\[3pt]
\multicolumn{2}{l}{Appendix} &17\\[3pt]
\multicolumn{2}{l}{Acknowledgements} &17
\end{tabular}}

\section{INTRODUCTION}
%\section{Introduction}
\label{sec:int}
Many years ago, Steven Weinberg mentioned to me that he was
disconcerted that none of the classic textbooks on
superconductivity would explain the phenomenon in terms of the
Higgs mechanism~\cite{sid} for the electromagnetic gauge field.  This
concern is of course very well justified, and it was most likely with
this concern in mind that Weinberg has included a section on
superconductivity in his treatment of spontaneous symmetry
breaking and the Higgs mechanism in his %by now likewise classic
series of volumes entitled {\it The quantum theory of
  fields}\hspace{2pt}~\cite{wein}.  When I was asked recently to
present some lectures on superconductivity, I opened his book
expecting to find a particularly lucid exposition of this in condensed
matter physics rarely emphasized perspective.  I found the exposition
I was looking for, but to my surprise, build around the
following statement: {\it A superconductor is simply a material in
  which electromagnetic gauge invariance is spontaneously broken.}
What Weinberg means with this statement is just that the
electromagnetic gauge field ``acquires a mass'' due to the Higgs
mechanism in a superconductor, as particle physicists often speak of
{spontaneously broken gauge invariance} interchangeably with
{the Higgs mechanism}.

Nonetheless, I am not perfectly at ease with the above statement,
which is, by the way, by no means specific to Weinberg's exposition,
but widely believed and accepted.  While it is obvious that Weinberg
fully understands the matter, the statement may still be misleading to
a young student who is learning the subject for the first time.  The
problem is that the statement is, if one takes it literally, not
correct: gauge invariance cannot spontaneously break down as a matter
of principle, and in particular is not broken in a superconductor, as
I will explain in the following section.

This paper is organized as follows.  In Section~\ref{sec:gau}, we
discuss the statement quoted above including the danger which may result
from a literal interpretation of it in depth.  In particular, we show
that the BCS ground state is, in contrast to statements made in the
literature, fully gauge invariant.  The crucial ingredient often
omitted is that gauge transformations involve, in addition to the
standard transformation of gauge fields, local phase rotations of {\it
  both} creation (and annihilation) operators {\it and} wave
functions.
In Section~\ref{sec:ord}, we discuss the nature of the order parameter in
superconductors, with a particular emphasis on finite systems, which
always possess a unique ground state.  The arising subtlety is
explained by drawing an analogy to quantum antiferromagnets, which
also possess a unique and rotationally invariant ground state for
finite systems.
In Section~\ref{sec:eff}, we motivate and elaborate the effective
field theory of a superconductor at low temperatures, which contains
the theory of a neutral superfluid as the special case where the
charge is set to zero.  In particular, we obtain the particle density
and current as well as the energy and momentum density from the
physical symmetries of the theory, invariance under global U(1) phase
rotations of the order parameter and invariance under translations in
time and space.  The quest for a consistent definition of the
superfluid velocity yields a relation between current and momentum
densities in the superfluid, which in turn requires corrections to the
effective Lagrangian.  Since the density of the superfluid is
essentially the ``momentum conjugate'' of the order parameter phase,
Hamilton's equations yield physical information not contained in the
Euler--Lagrange equations; specifically, we obtain a gauge invariant
generalization of Anderson's characteristic equations of superfluidity
to the case of superconductors.  We conclude this Section with a brief
justification of the classical nature of the effective field theory.
In Section~\ref{sec:phe}, we discuss the phenomenology of
superconductors as compared to neutral superfluids, or, in general
terms, the Higgs mechanism.  To begin with, we briefly address the
phenomenology of neutral superfluids including vortex quantization,
and
%review the formalism of 
give a general introduction to the Higgs mechanism in field theories.
We then turn to the phenomenology of simply connected superconductors,
solve the equations of motion, obtain electric and magnetic screening,
London's equation, the Bernoulli Hall effect, and the balance of the
Lorentz force.  We demonstrate that the Higgs mechanism never
corresponds to a spontaneous violation of a gauge symmetry, and that
it is incorrect to interpret it in terms of ``a mass acquired by the
electromagnetic gauge field'', as the massive field is no longer a
gauge field.  Specifically, we show that the massive vector field,
which may alternatively be used to describe a (simply connected)
superconductor, is correctly interpreted as a four-vector formed by
the chemical potential and the three components of the superfluid
velocity.  We conclude this Section with a discussion of the subtle
difference between the physical invariance of the theory under global
U(1) rotations of the order parameter phase and gauge invariance,
which is nothing but a local invariance of our description of the
system.
In the last Section, we review a family of ``quantum effects'': the
quantization of magnetic flux in superconductors, phase slippage, and
the Josephson effect in both neutral superfluids and superconductors.
In these effects, Planck's constant manifests itself in the
phenomenology through the compactness of the order parameter phase
field; these effects require either a non-trivial topology or more
than one superfluid.  We derive them from the effective field theory
introduced in section~\ref{sec:eff}, and thereby demonstrate that the
very few assumptions made in motivating the effective theory are
sufficient to account for them.

\section{GAUGE INVARIANCE}
%\section{Gauge invariance}
\label{sec:gau}

To illustrate how dangerous the statement quoted in the introduction
is in the case of superconductivity, where we do not only have a
description in terms of an effective field theory but also a
microscopic description in terms of model Hamiltonians and trial wave
functions, I will at first assume the statement was true and take it
literally.  I will pretend to be a student who has just learned that
electromagnetic gauge invariance is spontaneously violated in a
superconductor.  Well, what does this mean?  A spontaneously broken
symmetry means that the Hamiltonian of a given system in the
thermodynamic limit is invariant under a given symmetry transformation
(\ie commutes with the generator(s) of this symmetry) while the ground
state is not invariant.  There are many ground states, which transform
into each other under the symmetry transformations.  A classic example
is ferromagnetism: The Hamiltonian is rotationally invariant, while
any particular ground state, specified by the direction the
magnetization vector points to, is not.  So if gauge invariance is
broken in a superconductor, this must mean that the ground state of
the superconductor does not share the gauge invariance of the
Hamiltonian.  Indeed, a glance at the BCS wave
function~\cite{bcs,schrief,degenn,tinki}
%{BardeenCooperSchrieffer57pr1,BardeenCooperSchrieffer57pr2,Schrieffer1964,deGennes1966,Tinkham1996}
\begin{equation}
  \label{e:bcs}
  \ket{\psi%^{\scriptstyle\rm BCS}
    _\phi}=\prod_{\bs{k}}\left(u_{\bs{k}} + v_{\bs{k}} e^{i\phi} 
    c_{\bs{k}\up}^\dagger\,c_{-\bs{k}\dw}^\dagger\right)\vac,
\end{equation} 
where the coefficients $u_{\bs{k}}$ and $v_{\bs{k}}$ are chosen real
and $\phi$ is an arbitrary phase, appears to confirm this picture.
There are many different ground state wave functions, labeled by
$\phi$, which transform into each other under an electromagnetic gauge
transformation given by
\begin{equation}
  \label{e:gaugec}
  c_{\bs{k}\sigma}^\dagger\rightarrow 
  e^{i{\textstyle\frac{e}{\hbar c}}\Lambda} c_{\bs{k}\sigma}^\dagger,
\end{equation} 
which is tantamount to taking 
\begin{displaymath}
  \phi\rightarrow \phi+\frac{2e}{\hbar c}\Lambda .
\end{displaymath}
For simplicity, we have chosen $\Lambda$ independent of spacetime. 
The electron charge throughout this article is $-e$.

Next, I the student ask myself whether these many BCS wave functions
for different parameters $\phi$ correspond to physically different
states.  I know that gauge transformations are not physical
transformations: gauge invariance is an invariance of a description of
a system, while other symmetries correspond to invariances under
physical transformations, like rotations or translations, which affect
the physical state in question.  For example, if the Hamiltonian for
given system (like a ferromagnet) is invariant under rotations in
space, this implies that if we rotate a given eigenstate, we will
obtain another eigenstate.  Depending on whether the original state is
rotationally invariant or not, it will transform into itself or into a
physically different state.  A gauge transformation, by contrast, will
only transform our description of a system from one gauge to another,
without ever having any effect on the physical state of the system.
Gauge transformations are comparable to rotations or translations of
the coordinate system we use to describe a system.  Another way of
seeing the difference is by noting that it is possible to rotate or
translate a superconductor in the laboratory, but as a matter of
principle not possible to gauge transform it.  Returning to the
superconductor, I the student conclude that if the many different
ground states only differ by a gauge transformation, they cannot be
physically different.  The ground state of a superconductor must hence
be physically unique.

In fact, there is another way of looking at the problem which appears
to confirm this conclusion.  A BCS superconductor can not only be
described in the grand-canonical ensemble, where the chemical
potential rather than the number of particles is fixed, but also in
the canonical ensemble, where the number of particles or pairs is
fixed.  Following Anderson~\cite{phil66}, we can project out a (not
normalized) state with $N$ pairs from (\ref{e:bcs}) via
\begin{equation}
  \label{e:project}
  \ket{\psi_N} = \int_0^{2\pi}\!\!d\phi\, e^{-iN\phi}\ket{\psi_\phi}
\end{equation} 
and obtain (see appendix)
\begin{eqnarray}
  \label{e:bcsn}
  \ket{\psi_N}\!\!\! &=&\!\!\!\int\!\! 
  d^{3\,}\!\bs{x}_1\!\ldots d^{3\,}\!\bs{x}_{2\!N}\,
  \varphi(\bs{x}_1-\bs{x}_2)\cdot\!\ldots\!\cdot
  \varphi(\bs{x}_{2\!N-1}\!-\bs{x}_{2\!N})
  \,\cdot\cr
  && \cdot\,\psi^\dagger_\up(\bs{x}_1)\psi^\dagger_\dw(\bs{x}_2)\ldots
  \psi^\dagger_\up(\bs{x}_{2N-1})\psi^\dagger_\dw(\bs{x}_{2N})\vac \!,
\end{eqnarray}
where the real-space creation operator fields $\psi^\dagger_\sigma(\bs{x})$
are simply the Fourier transforms of the momentum-space creation operators
$c_{\bs{k}\sigma}^\dagger$,
\begin{equation}
  \label{e:ft}
  \psi^\dagger_\sigma(\bs{x})=\frac{1}{\sqrt{V}}\sum_{\bs{k}} 
  e^{-i\bs{k}\bs{x}} c^\dagger_{\bs{k}\sigma }, \
  c^\dagger_{\bs{k} \sigma}=\frac{1}{\sqrt{V}}\!\int\!\! d^{3\,}\!\bs{x}\, 
  e^{i\bs{k}\bs{x}} \psi^\dagger_\sigma(\bs{x}).
\end{equation}
% \end{displaymath}
The wave function for each of the individual pairs, which only depends
on the relative coordinate, is (up to a normalization) given by
\begin{equation}
  \label{e:varphi}
  \varphi(\bs{x})=\frac{1}{V}\sum_{\bs{k}} \frac{v_{\bs{k}}}{u_{\bs{k}}}
  e^{i\bs{k}\bs{x}}.  
\end{equation}
This form nicely illustrates that all the
pairs have condensed into the same state, which is the essence of
superfluidity.  As $\varphi(\bs{x})$ is uniquely determined for a given
Hamiltonian, the ground state of the superconductor (\ref{e:bcsn})
once more appears to be unique and non-degenerate.

So far the students train of thought.  The conclusion reached is of
course completely wrong:  a superfluid, and in particular a superconductor,
is characterized by a spontaneously broken symmetry, and, at least in the
thermodynamic limit, there are many degenerate ground states.  There 
are several mistakes in the students analysis.  The first
is his literal interpretation of the statement quoted in the
inroduction.  In fact, {\it a gauge symmetry cannot spontaneously
  break down as a matter of principle, since it is not a physical
  symmetry of the system to begin with, but merely an invariance of
  description}~\cite{elit}. The only way to violate a gauge symmetry is
by choosing a gauge, which again has only an effect on our
description, but not on the physical system itself.

In particular, the BCS ground state does not violate gauge invariance,
even though statements to the contrary have been made in the literature.
The apparent contradiction with (\ref{e:bcs}) and 
(\ref{e:gaugec}) can be resolved by recalling that a gauge transformation
only affects our description of the system, and is analogous to a rotation
of the coordinate system we use in the example of a ferromagnet:  if
we rotate the coordinate system accordingly, a ground state with 
the magnetization vector pointing in the $z$-direction in the original 
coordinate system will ``transform'' into a state with the magnetization
vector pointing in the $x$-direction in the new coordinate system,
while the physical state has not been affected at all.
So while the BCS wave function may look different in a different gauge,
the state itself will remain the same.

It is worthwhile to rephrase this statement in equations.
To begin with, let us consider a (relativistic quantum) field
theory.  Electromagnetic gauge invariance is the invariance of a given
theory under U(1) rotations of the complex scalar fields which carry
the charge:
\begin{equation}
  \label{e:gaugeprel}
  \psi^\dagger(x)\rightarrow 
  e^{i{\textstyle\frac{e}{\hbar c}}\Lambda(x)}\psi^\dagger(x),\quad 
  \psi(x)\rightarrow e^{-i{\textstyle\frac{e}{\hbar c}}\Lambda(x)}\psi(x),
\end{equation}
where $x$ denotes spacetime.  We use the conventions
$(x^\mu)=(x^0,x^1,x^2,x^3)=(ct,x,y,z)$,
$(x_\mu)=g_{\mu\nu}x^\nu$, $1=g_{00}=-g_{11}=-g_{22}=-g_{33}$.
If the theory contains gradient terms in these fields (as it usually does),
gauge invariance demands that they are minimally coupled to 
a U(1) gauge field, \ie the gradient terms must enter the Lagrangian as 
\begin{displaymath}
  \Bigl(\partial_\mu + i{\frac{e}{\hbar c}} A_\mu (x)\Bigr)\psi^\dagger(x)
  \quad \text{or}\quad
  \Bigl(\partial_\mu - i{\frac{e}{\hbar c}} A_\mu (x)\Bigr)\psi(x),
\end{displaymath}
where $(\partial_\mu)\equiv (\partial/\partial x^\mu) =
({\frac{1}{c}}\partial_t, \nabla)$.  The gauge field
$(A_\mu)=(\Phi,-\bs{A})$ must transform according to
\begin{equation}
  \label{e:gaugearel}
  A_\mu(x)\rightarrow A_\mu(x)-\partial_\mu \Lambda(x).  
\end{equation}
The statement that the theory is gauge invariant simply means
that the Lagrangian is invariant under the combined transformation
(\ref{e:gaugeprel}) and (\ref{e:gaugearel}).  It is not a physical
invariance, but an invariance of description, as it only amounts to a
reparametrization of fields. 

The concept of gauge invariance is implemented in a very similar way
in non-relativistic quantum mechanics, where the gauge field $\bs{A}$ 
is no longer considered a dynamical variable, but an externally applied 
vector potential, and we usually do not describe a system by
a Lagrange density, but by a Hamiltonian operator and its
eigenstates.  For pedagogical reasons, let us first assume a formulation 
in second quantization.  Electromagnetic gauge invariance means once
again that the description is invariant under U(1) rotations of the
particle creation and annihilation operator fields~\cite{simp},  
\begin{equation}
  \label{e:gaugep}
  \psi^\dagger_\sigma(\bs{x})\rightarrow 
  e^{i{\textstyle\frac{e}{\hbar c}}\Lambda(\bs{x})}
  \psi_\sigma^\dagger(\bs{x}), \quad 
  \psi_\sigma(\bs{x})\rightarrow
  e^{-i{\textstyle\frac{e}{\hbar c}}\Lambda(\bs{x})}\psi_\sigma(\bs{x}).
\end{equation}
The kinetic part of the Hamiltonian will again contain gradient terms
in the operator fields, which once again must be minimally coupled to
the electromagnetic gauge field.  For example, the standard kinetic
Hamiltonian for a quadratic dispersion 
\begin{equation}
  \label{e:ham}
  H_{\scriptscriptstyle\textrm{kin}}=
  \frac{1}{2m}\sum_\sigma\!\int\!d^{3\,}\!\bs{x}\,
  \psi^\dagger_\sigma(\bs{x}) 
  \left(-i\hbar\nabla +i\frac{e}{\hbar}\bs{A(\bs{x})}\right)^2
  \psi_\sigma(\bs{x})
\end{equation}
is obviously invariant under (\ref{e:gaugep}) provided we transform
the gauge field simultaneously according to
\begin{equation}
  \label{e:gaugea}
  \bs{A(\bs{x})}\rightarrow \bs{A(\bs{x})}+\nabla\Lambda(\bs{x}).
\end{equation}
Let us now turn to the gauge transformation properties of the
eigenstates.  Consider a general $N$ electron eigenstate
\begin{eqnarray}
  \label{e:eigen}
  \ket{\varphi}\!\!\! &=&\!\!\!\sum_{\sigma_1\ldots\sigma_N}\!\int\!
  d^{3\,}\!\bs{x}_1\ldots d^{3\,}\!\bs{x}_N\,
  \varphi(\bs{x}_1\ldots \bs{x}_N;\sigma_1\ldots\sigma_N)\,\cdot\cr
  &&\rule{0pt}{12pt}\hspace{15mm}
  \psi^\dagger_{\sigma_1}(\bs{x}_1)\ldots\psi^\dagger_{\sigma_N}(\bs{x}_N)
  \vac \!.
\end{eqnarray}
The state is invariant under (\ref{e:gaugep}) provided we transform
the wave function according to
\begin{eqnarray}
  \label{e:gaugepsi}
  &&\hspace{-9mm}
  \varphi(\bs{x}_1\ldots \bs{x}_N;\sigma_1\ldots\sigma_N)\,\rightarrow\,\cr
  &&\rule{0pt}{22pt}\hspace{1mm}
  \prod_{j=1}^N e^{-i{\textstyle\frac{e}{\hbar c}}\Lambda(\bs{x_j})}\,
  \varphi(\bs{x}_1\ldots \bs{x}_N;\sigma_1\ldots\sigma_N).
\end{eqnarray}
This already illustrates the statement phrased in words above: A gauge
transformation leaves physical states invariant.  This is just not
obvious in every formulation.  If we formulate a problem in
non-relativistic quantum mechanics in first quantization, a gauge
transformation will only amount to (\ref{e:gaugea}) and
(\ref{e:gaugepsi}), as we do not even introduce the operator fields
$\psi^\dagger$ and $\psi$.  As $\bs{A}(\bs{x})$ implements an
externally applied magnetic field, we must choose a gauge in order to
obtain explicit expressions for the Hamiltonian and the eigenstates.
The vector potential and the wave functions will have different
functional forms in different gauges.  The gauge rotations
(\ref{e:gaugep}) of the particle creation and annihilation operators,
by contrast, only amount to a local change of variables; we could write
\begin{displaymath}
  \begin{array}{rcl}\psi^\dagger_\sigma(\bs{x}) \!\!&\! \rightarrow \!&\!\!
    {\psi^\dagger_\sigma}'(\bs{x})=
    e^{i{\textstyle\frac{e}{\hbar c}}\Lambda(\bs{x})}\,
    \psi_\sigma^\dagger(\bs{x}),\\\rule{0pt}{18pt}
    \quad \psi_\sigma(\bs{x}) \!\!&\! \rightarrow \!&\!\!
    {\psi_\sigma}'(\bs{x})= 
    e^{-i{\textstyle\frac{e}{\hbar c}}\Lambda(\bs{x})}\,
    \psi_\sigma(\bs{x}),
\end{array}
\end{displaymath}
and then simply omit the primes.  This part of the gauge
transformation is often omitted as a choice of convention.

In the case of a BCS superconductor, such a convention would be all but
propitious, as it would suggests that the ground state is not gauge
invariant.  The apparent contradiction in the students train of
thought is immediately resolved as one uses the full and correct
prescription for a gauge transformation, 
\begin{equation}
  \label{e:gaugecc}
  c_{\bs{k}\sigma}^\dagger\rightarrow 
  e^{i{\textstyle\frac{e}{\hbar c}}\Lambda} c_{\bs{k}\sigma}^\dagger,
  \qquad \phi\rightarrow\phi-\frac{2e}{\hbar c}\Lambda,
\end{equation} 
where the transformation of the phase $\phi$ is the equivalent of 
(\ref{e:gaugepsi}) above.  Then the BCS ground state 
\begin{displaymath}
  \prod_{\bs{k}}\left(u_{\bs{k}} + v_{\bs{k}} e^{i\phi} 
    c_{\bs{k}\up}^\dagger\,c_{-\bs{k}\dw}^\dagger\right)\vac \!,
\end{displaymath}
is evidently gauge invariant; it is merely the label $\phi$ in
$\ket{\psi_\phi}$ which will be adjusted under a gauge transformation.
The transformation $\phi\rightarrow\phi-\frac{2e}{\hbar c}\Lambda$ is
also required for the classical (or Ginzburg--Landau) order parameter
field $\Psi^*(\bs{x})$, which is given by the expectation value of the
operator field
\begin{equation}
  \label{e:op}
  \hat\Psi^\dagger(\bs{x})\equiv 
  \psi^\dagger_\up(\bs{x})\,\psi^\dagger_\dw(\bs{x}),
\end{equation}
to have to the correct gauge transformation properties.  
For the BCS ground state,
\begin{equation}
  \label{e:bcsop}
  \Psi^*(\bs{x})= %\propto %\equiv
  \bra{\psi_\phi}\,\psi^\dagger_\up(\bs{x})\,\psi^\dagger_\dw(\bs{x}) 
  \ket{\psi_\phi}
  =\frac{1}{V}\sum_{\bs{k}} v_{\bs{k}}^* u_{\bs{k}}^{\phantom{*}}e^{-i\phi}
\end{equation} 
will transform as a field of charge $-2e$ under (\ref{e:gaugecc}),
\begin{equation}
  \label{e:opgauge}
  \Psi^*(\bs{x})\rightarrow 
  e^{i{\textstyle\frac{2e}{\hbar c}}\Lambda(\bs{x})}\Psi^*(\bs{x}).
\end{equation} 
This is the physically correct prescription.  When we couple
$\Psi^*(\bs{x})$ minimally to the electromagnetic gauge field, as
required by (\ref{e:opgauge}), we obtain the correct effective field
theory description of superconductivity.  This theory displays the
Higgs mechanism and yields London's equation.  (By contrast, if we
were to adhere to (\ref{e:gaugec}), $\Psi^*(\bs{x})$ would be
invariant, could not be coupled to the electromagnetic gauge field,
and no sensible effective field theory could be formulated.)

\section{ORDER PARAMETER CONSIDERATIONS}
%\section{Order parameter considerations}
\label{sec:ord}

Before proceeding further with the Higgs mechanism, %in this direction, 
I would like to return to the students train of thought and explain
what is wrong with his conclusion drawn from the BCS wave function in
position space.  The problem here is that while the ground state is
indeed unique for a finite system, there are many degenerate states in
the thermodynamic limit, which correspond to different numbers of
particles.  To understand this issue in depth, it is best to first
recall how rotational symmetry is spontaneously violated in
ferromagnets and antiferromagnets.  As a minimal model, we consider a
three dimensional cubic lattice of spins with spin quantum number $S$
and assume the Heisenberg Hamiltonian~\cite{auer}
\begin{equation}
  \label{e:heisen}
  H_J = J \sum_{\bond} \bs{S}_i \bs{S}_j
\end{equation} 
where the sum extends over all nearest-neighbor bonds $\bond$ and
$J<0$ ($J>0$) for a ferromagnet (an antiferromagnet).

In the case of a ferromagnet, the order parameter is given
by the total spin operator 
\begin{equation}
  \label{e:stot}
  \bs{S}_{\rm tot}\equiv\sum_i \bs{S}_i,
\end{equation} 
where the sum extends over all lattice sites.  It commutes with the
Hamiltonian,
\begin{equation}
  \label{e:commstot}
  \comm{H_J}{\bs{S}_{\rm tot}}=0,
\end{equation}
and it is hence possible to choose simultaneous eigenstates of the
Hamiltonian and the order parameter.  In other words, the degenerate
eigenstates of the order parameter corresponding to all possible
directions the magnetization vector
\begin{displaymath}
  \bs{M}=\braket{\bs{S}_{\rm tot}}%_0
\end{displaymath}
can point to, are simultaneously degenerate eigenstates of the
Hamiltonian.  (For a ferromagnet with $N$ spins, all the spins align
and the ground states are just the states with maximal total spin
$S_{\rm tot}=NS$.)  The ground state of the Hamiltonian is vastly
(\ie $2S_{\rm tot}\!+\!1$ fold) degenerate even if the
system is finite.

\begin{figure}[tbhp]
  \begin{center}
    \vspace{2mm}
    \includegraphics[width=\columnwidth]{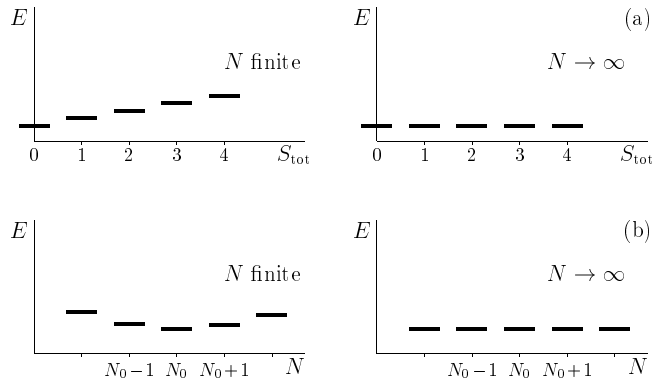}
  \end{center}
  \caption{In antiferromagnets (a) and superconductors (b), the 
    ground state is unique for finite systems but degenerate in the
    thermodynamic limit.}
  \label{f:spec}
\end{figure}

The situation is different in the case of the antiferromagnet.  The
order parameter is given by the N\'eel vector, which in operator form
is given by
\begin{equation}
  \label{e:neel}
  \bs{\hat N}\equiv\sum_{i\in {\rm A}}\bs{S}_i-\sum_{j\in {\rm B}}\bs{S}_j,
\end{equation} 
where A and B denote the two sublattices of the (bipartite) cubic
lattice.  It does {\it not} commute with the Hamiltonian:
\begin{displaymath}
  \comm{H_J}{\bs{\hat N}}\ne 0.
\end{displaymath}
This implies that we cannot choose simultaneous eigenstates for the
Hamiltonian and the order parameter.  In fact, a theorem due to
Marshall~\cite{marshall} states that the ground state for $N$ even is
unique and a spin singlet, or in other words, rotationally invariant.
(It is possible to choose simultaneous eigenstates of $H_J$ and
$\bs{S}_{\rm tot}$ as (\ref{e:commstot}) holds independently of the
sign of $J$.)  The classical N\'eel order parameter,
\begin{displaymath}
  \bs{N}=\bigl\langle\bs{\hat N}\bigr\rangle%_0
\end{displaymath}
will vanish for any finite system.  This is not to say that there is
no order for a finite system; it just manifests itself only through
long-range correlations in the staggered spin-spin correlation function: 
\begin{displaymath}
  \braket{\bs{S}_i \bs{S}_j }\rightarrow \pm\ \hbox{const.}\qquad 
  \hbox{as}\ i\!-\!j\rightarrow\infty
\end{displaymath}
where the $+$ sign applies for $i$ and $j$ on the same sublattice, 
the $-$ sign for $i$ and $j$ on different sublattices, 
and $i\!-\!j\rightarrow\infty$ is understood 
to denote a very large separation within (the finite volume of) the 
system.  As we approach the thermodynamic limit,
the difference in energy between the lowest singlet and lowest
eigenstates for %higher total spin 
$S_{\rm tot}=1,2,3\ldots$ vanishes, and the ground state becomes
degenerate (see Fig.\ \ref{f:spec}a).  These degenerate states can
now be classified by the directions of the N\'eel vector $\bs{N}$, and
the spontaneous breakdown of rotation symmetry is evident.

The situation in superconductors is analogous to the antiferromagnet:
The (operator valued) order parameter (\ref{e:op}) does not commute
with the BCS Hamiltonian for any finite system even if we work in the
grand-canonical ensemble, and the ground state for any finite volume
will have a well defined particle number.  The difference in energy
between a system with $N$ or $N\pm 1$ or $N\pm 2$ {\it etc.}\ pairs,
however, will vanish in the thermodynamic limit (see
\mbox{Fig.~\ref{f:spec}b)}, and the many degenerate ground states can
be classified by the phase $\phi$ of the (classical) order parameter
\begin{equation}
  \label{e:bcsop2}
  \Psi^*(\bs{x})= \braket{\hat\Psi^\dagger(\bs{x})}=
  |\Psi^*(\bs{x})| e^{-i\phi(\bs{x})}.
\end{equation} 
The broken symmetry is of course also present in a system with a fixed
number of particles, but like in the case of the antiferromagnet, only
as a long-range correlation of the (operator valued) order parameter
field:
\begin{equation}
  \label{e:odlro}
  \braket{\hat\Psi^\dagger(\bs{x})\hat\Psi(\bs{y})}\rightarrow 
  \hbox{const.}\qquad \hbox{as}\ |\bs{x}-\bs{y}|\rightarrow\infty,
\end{equation}
where $\hat\Psi(\bs{x})\equiv \psi_\dw(\bs{x})\psi_\up(\bs{x})$
is simply the hermitian conjugate of $\hat\Psi^\dagger(\bs{x})$.  This
correlation is referred to as off diagonal long-range order
(ODLRO)~\cite{odlro}.  This type of order is characteristic to all
superfluids, whether charged (like a superconductor) or neutral (like
liquid helium), whether fermionic (like a superconductor or $^3$He) or
bosonic (like $^4$He).  For a bosonic superfluid, the (operator
valued) order parameter $\hat\Psi^\dagger(\bs{x})$ and its hermitian
conjugate $\hat\Psi(\bs{x})$ no longer create or annihilate a pair of
fermions, but simply create or annihilate a single boson (like a
$^4$He atom).

The ODLRO is already evident from the position space wave function
(\ref{e:bcsn}): Since all the pairs have condensed into the same
quantum state, which is translationally invariant as it does not
depend on the center-of-mass coordinates of the pairs, we expect to
obtain a finite overlap with the original ground state if we rather
clumsily (\ie via $\hat\Psi(\bs{y})$) remove a pair of particles at some
location $\bs{y}$ and equally clumsily (\ie via
$\hat\Psi^\dagger(\bs{x})$) recreate it at a distant location
$\bs{x}$.  In a superfluid or superconductor with a fixed number of
particles, the phase $\phi$ will align over the entire system, like
the direction of the staggered magnetization or N\'eel vector will
align in an antiferromagnet.

To illustrate the significance of the phase once more, let us consider
a large (but finite) superconductor A, and describe it as a
combination of two superconductors B and C: %(see \mbox{Fig.~\ref{fig:abc})}.  
\begin{center}
%\begin{picture}(180,40)(-100,-20)
\begin{picture}(180,40)(-90,-20)
% horizonal lines
\put(-55,-16){\makebox(0,0){\rule{70.pt}{ 0.3pt}}}
\put( 55,-16){\makebox(0,0){\rule{70.pt}{ 0.3pt}}}
\put(-55, 16){\makebox(0,0){\rule{70.pt}{ 0.3pt}}}
\put( 55, 16){\makebox(0,0){\rule{70.pt}{ 0.3pt}}}
%vertical lines
\put(-90,0){\makebox(0,0){\rule{ 0.3pt}{32.pt}}}
\put(-20,0){\makebox(0,0){\rule{ 0.3pt}{32.pt}}}
\put( 20,0){\makebox(0,0){\rule{ 0.3pt}{32.pt}}}
\put( 62,0){\makebox(0,0){\rule{ 0.3pt}{32.pt}}}
\put( 90,0){\makebox(0,0){\rule{ 0.3pt}{32.pt}}}
%labels
\put(  0,  0){\makebox(0,0){\large $=$}}
\put(-55,  0){\makebox(0,0){\large A}}
\put( 41,  0){\makebox(0,0){\large B}}
\put( 76,  0){\makebox(0,0){\large C}}
\end{picture}
\end{center}
If we label the ground states of each superconductor by its phase, 
we can obviously write
\begin{displaymath}
  \ket{\psi_\phi^{\text{A}}}=
  \ket{\psi_\phi^{\text{B}}}\otimes\ket{\psi_\phi^{\text{C}}}
\end{displaymath}
as the phase $\phi$ of the order parameter will align over the entire system.
If we now transform to a description in terms of fixed numbers of
pairs $N_a$ for each superconductor,
\begin{displaymath}
  \ket{\psi^a_N} = \int_0^{2\pi}\!\!d\phi\, 
  e^{-iN_a\phi}\ket{\psi^a_\phi},
\end{displaymath}
where $a$ can be A, B, or C, the ground state of A is no longer
a direct product of the ground states of B and C:
\begin{displaymath}
  \ket{\psi_{N_{\text{A}}}^{\text{A}}}
  \ne\ket{\psi_{N_{\text{B}}}^{\text{B}}}
  \otimes\ket{\psi_{N_{\text{A}}-N_{\text{B}}}^{\text{C}}},
\end{displaymath}
no matter how we choose $N_{\text{B}}$, as the phases no longer align.  So
while it is possible to describe a superconductor in a canonical
ensemble (\ie with fixed particle number), it is highly awkward to do
so.  It is comparable to a description of an antiferromagnet with
long-range N\'eel order in terms of an overall singlet ground state of
the system.

The most significant difference between an antiferromagnet and a
superfluid or superconductor with regard to %the concept of 
the order parameter is that the broken rotational symmetry in the
former case is much more evident to us, as all the macroscopic objects
in our daily life experience violate rotational symmetry at one level
or another.  In particular, the structure of the material in which
antiferromagnetic order occurs provides us already with a reference
frame for the direction the N\'eel order parameter may point to.  In
the case of the superconductor, we need a second superconductor to
have a reference direction for the phase, and an interaction between
the order parameter in both superconductors to detect a relative
difference in the phases.  (In practice, such an interaction may be
accomplished by a pair tunneling or so-called Josephson junction.)
The interference experiments will of course only be sensitive to the
relative phase, and not the absolute phase in any of the
superconductors, as all phases can, as a matter of principle, only be
specified relative to some reference phase.  In principle, the same is
true for rotational invariance, but in this case the fixed stars
provide us with a reference frame we perceive as ``absolute''.

We may conclude at this point that in a superfluid or superconductor,
a symmetry is spontaneously violated, but this symmetry is {\em not}
gauge invariance, but global U(1) phase rotation symmetry. This is
already evident from the fact that the discussion above made no
reference to whether the order parameter field
$\hat\Psi^\dagger(\bs{x})$ is charged or not, and equally well applies
to neutral superfluids, where $\hat\Psi^\dagger(\bs{x})$ carries no
charge.

There is, however, a very important difference between these two
cases.  If the order parameter field is neutral, the excitation
spectrum of the system contains a gapless (or in the language of
particle physics ``massless'') mode, a so-called Goldstone
boson~\cite{sid}, which physically corresponds to very slow spatial
variations in the direction (as for the case of broken rotational
invariance) or phase (as for the case of a superfluid) of the
classical order parameter field.  If the order parameter field is
charged, however, it couples to the electromagnetic gauge field, and
the Goldstone boson is absent due to the Higgs mechanism.  The
physical principle underlying the mechanism was discovered by
Anderson~\cite{phil63} in the context of superconductivity: as the
electromagnetic interaction is long-ranged, the mode corresponding to
very slow spatial variations in the phase $\phi$ of the
superconducting order parameter, which implies currents by the
equation of motion and hence also variations in the density of the
superfluid by the continuity equation, acquires a gap (or ``mass'')
given by the plasma frequency.

\section{EFFECTIVE FIELD THEORY} 
%\section{Effective field theory} 
\label{sec:eff}

Most of the phenomenology of superfluidity or superconductivity can be
derived from a simple effective field theory, which in the latter
case displays the Higgs mechanism.
It is probably best to turn directly to the low-energy effective
Lagrangian for the superconductor, as it contains the superfluid as
the special case where the coupling $e^*$ of the order parameter to
the electromagnetic gauge field is set to zero.  To motivate the
Lagrangian, recall first the Ginzburg--Landau~\cite{lg} expansion of the
free energy density in terms of the order parameter (which is now
normalized differently from (\ref{e:bcsop}) above) in the vicinity of
the critical temperature $T_{\text{c}}$, where the transition between normal
and superconducting phases occurs:
\begin{eqnarray}
  f(T,\Psi)\!\!&\!=\!&\!\! 
  \frac{1}{2m^*} \biggl|\Bigl(
  -i\hbar\nabla+\frac{e^*}{c}\bs{A}(\bs{x})\Bigr)\Psi(\bs{x})\biggr|^2
  \nonumber \\ \label{e:lg} \rule{0pt}{20pt} 
  &&\hspace{-10mm}+\, a(T)|\Psi(\bs{x})|^2+\frac{1}{2}b(T)|\Psi(\bs{x})|^4
  + \frac{1}{8\pi}\bs{B}(\bs{x})^2\hspace{5mm}
\end{eqnarray}
where $m^*$ and $-e^*=-2e$ are the effective mass and charge of the
electron pairs, respectively, and
$\bs{B}=\nabla\times\bs{A}$ is the magnetic field.
The material parameter $a(T)$ changes sign to become negative as we
pass through the transition from above, while $b(T)$ has to remain
positive.  
Minimizing the free energy in the superconducting phase yields that
(i) the gradient term must vanish,
(ii) $|\Psi(\bs{x})|^2=-{a}/{b}$,
%\begin{displaymath}
%  |\Psi(\bs{x})|^2=-\frac{a}{b}\, ,
%\end{displaymath}
and (iii) $\bs{B}=0$.  
This means that the amplitude $\Psi_0$ of the order parameter
$\Psi(\bs{x})=\Psi_0e^{i\phi}$ has to be fixed while the phase $\phi$,
which labels the many degenerate ground state configurations, can be
arbitrary as long as the variation over the sample is given by
\begin{displaymath}
  \nabla\phi=-\frac{e^*}{\hbar c}\bs{A},
\end{displaymath}
which implies $\phi(\bs{x})=\text{const.}$ if we choose the gauge
$\bs{A}(\bs{x})=0$.  In the vicinity of the transition, we may treat
$\Psi$ as a small parameter, which implies that the expansion
(\ref{e:lg}) provides us with a complete description of the system at
the level of thermodynamics.

The Ginzburg--Landau expansion is also helpful in motivating the low
energy effective Lagrange density at low temperatures.  To begin with, 
we may assume that since the amplitude fluctuations are massive, they
do not enter in the low energy description.
Taking $|\Psi(\bs{x})|$ to be constant, the free energy density above
reduces to a constant, an electromagnetic field contribution, and
\begin{equation}
  \label{e:fmag}
  f_{\text{mag}}=\frac{n_{\text s}}{2m^*}\,
  \Big(\hbar\nabla\phi(\bs{x})+\frac{e^*}{c}\bs{A}(\bs{x})\Bigl)^2,
\end{equation}
where $n_{\text s}=|\Psi_0|^2$ is a phenomenological parameter which
depends on the material and the temperature.  It has the dimension of
a density and is equal to the density of the superfluid in the absence
of currents and inhomogeneities at $T=0$, as we shall see below.  It
is usually referred to as the superfluid density, but it would be more
appropriate to use the superfluid stiffness ${n_{\text s}}/{m^*}$ as a
parameter instead~\cite{phil66}.  We will also see below that Galilean
invariance of the superfluid implies that $m^*$ is the bare mass of
the superfluid particles, \ie $m^*=2m_{\text{e}}$ for Cooper
pairs~\cite{rel}.

We take (\ref{e:fmag}) to be part of the potential energy in the
effective Lagrange density for the superfluid.  The remaining
contribution arises from the coupling of the charge of the superfluid
to the electrostatic potential $\Phi(\bs{x})$, which is in leading
order given by
\begin{displaymath}
  f_{\text{el}}=-n_{\text s} e^*\Phi(\bs{x}).
\end{displaymath}
This term is usually not included in the free energy of the
superconductor, as it is always canceled off by another such term with
opposite sign arising from the uniform positive background charge.  It
is essential to our effective field theory here, however, as it is
part of the Lagrange density for the superfluid, while the uniform
background charge is accounted for by another Lagrange density
\begin{equation}
  \label{e:lb}
  \mathcal{L}_{\text b}(x)=-n_{\text{s}} e^*\Phi(x),
\end{equation}
where $x=(ct,\bs{x})$ denotes spacetime.  Note that $f_{\text{el}}$
is not invariant under (time dependent) gauge transformations
\begin{eqnarray}
  \label{e:gauge}
  \phi(x)\!&\!\to \!&\!\phi(x)-\frac{e^*}{\hbar c}\Lambda(x),
  \nonumber\\\rule{0pt}{16pt}
  \Phi(x)\!&\!\to \!&\!\Phi(x)-\frac{1}{c}\partial_t \Lambda(x),
  \\\rule{0pt}{12pt}
  \bs{A}(x)\!&\!\to \!&\!\bs{A}(x)+\nabla\Lambda(x).
  \nonumber
\end{eqnarray}

We now turn to the kinetic energy term in the effective Lagrange density.
The simplest gauge invariant Lagrange density containing
both potential energy terms above is given by~\cite{feyn}
\begin{eqnarray}
\label{e:lmin}
\mathcal{L}_{\text s}(x)
\!&\!=\!&\!-\,n_{\text{s}} 
\Big(\hbar\partial_t \phi(x)-e^*\Phi(x)\Bigl)\cr\rule{0pt}{20pt}
&&\!-\,\frac{n_{\text{s}}}{2m^*}
\Big(\hbar\nabla\phi(x)+\frac{e^*}{c}\bs{A}(x)\Bigl)^2.\hspace{5mm}
\end{eqnarray}
This Lagrange density, however, cannot be complete.  The only term
containing a time derivative, $\partial_t \phi(x)$, appears as a total
derivative (here time derivative of $\phi$) and hence does not affect
the Euler--Lagrange equations of motion.  It is nonetheless of
physical significance, as it both ensures gauge invariance and
accounts for the leading contribution to the particle density, as we
will see below.

To obtain a second order time derivative term, 
recall that the characteristic feature of a neutral (\ie $e^*=0$)
superfluid is that the only excitation at low energies is a sound wave
with a linear dispersion
\begin{equation}
\label{e:lindis}
\omega (\bs{k}) =v|\bs{k}|,
\end{equation}
where $\bs{k}$ is the wave number and $v$ is the velocity of sound in
the fluid.  As we wish the effective Lagrange density for the
superfluid both to be gauge invariant and to yield (\ref{e:lindis}) as
an equation of motion for $e^*=0$, we arrive at
\begin{eqnarray}
  \label{e:ls}
  \mathcal{L}_{\text s}(x)
  \!&\!=\!&\!-\,
  n_{\text{s}}\Big(\hbar\partial_t \phi(x)-e^*\Phi(x)\Bigl)
  \\ \nonumber\rule{0pt}{18pt}
  &&\hspace{-13mm}\!\!+\frac{n_{\text{s}}}{2m^*}
  \biggl\{\frac{1}{v^2}\Big(\hbar\partial_t \phi(x)\!-\!
  e^*\Phi(x)\Bigl)^2- 
  \Big(\hbar\nabla\phi(x)\!+\!\frac{e^*}{c}\bs{A}(x)\Bigl)^{\!2}
  \biggr\}.
\end{eqnarray}
With 
\begin{equation}
  \label{e:dmu}
  D_\mu\phi\equiv\hbar\:\! \partial_\mu \phi-\frac{e^*}{c}A_\mu\:\!, 
\end{equation}
where $(\partial_\mu) %\equiv\frac{\partial}{\partial x^\mu}
= ({\frac{1}{c}}\partial_t, \nabla)$ and $(A_\mu)=(\Phi,-\bs{A})$,
the Lagrange density may also be written
\begin{equation}
  \label{e:ls2}
  \mathcal{L}_{\text s}=-%\:\!\hbar
  \:\! c\:\! n_{\text{s}}\;\!D_0\phi\:\!+\:\!
  \frac{n_{\text{s}}}{2m^*}\,
  \biggl\{\frac{c^2}{v^2}\,(D_0\phi)^2-(D_i\phi)^2\biggr\},
\end{equation}
where $i=1,2,3$.
The total Lagrangian of the system is given by
\begin{equation}
  \label{e:ltot}
  L=\int\! d^{3\,}\!\bs{x} 
  \left\{\mathcal{L}_{\text s}(x) + \mathcal{L}_{\text b}(x) + 
    \mathcal{L}_{\text{em}}(x)\right\},
\end{equation}
where 
\begin{equation}
  \label{e:lmax}
  \mathcal{L}_{\text{em}}=-\frac{1}{16\pi}F_{\mu\nu}F^{\mu\nu}
  \quad\text{with}\quad
  F_{\mu\nu}\equiv\partial_\mu A_\nu-\partial_\nu A_\mu
\end{equation}
denotes the standard Maxwell Lagrange density for electromagnetism. 

The astonishing feature is now that this simple Lagrangian for the
compact U(1) field $\phi(x)$ (compact since the values $\phi$ and
$\phi+2\pi$ describe the same physical state and hence must be
identified) coupled to the electromagnetic gauge field accounts for
all the essential features of superfluidity or superconductivity.
There are also important corrections to it, but we will discover
them automatically as we proceed.

In order understand the physical content of the Lagrangian, it is
highly instructive to study its symmetries, in particular particle
number conservation and invariance under translations in space and
time.  We wish our analysis to apply both to the case of a neutral and
a charged superfluid.  In the former case, the theory is no longer
invariant under a local U(1) gauge transformation (as the
electromagnetic gauge transformation (\ref{e:gaugeprel}) reduces to
the identity transformation for $e^*=0$) but still invariant under a
global U(1) rotation
\begin{equation}
  \label{e:rot}
  \Psi(x)\rightarrow 
  e^{i\lambda}\Psi(x)\quad\text{or}\quad
  \phi(x)\rightarrow \phi(x)+\lambda,
\end{equation}
where $\lambda$ is independent of spacetime.  Physically, this
symmetry corresponds to particle (or Cooper pair) number conservation.
According to Noether's theorem~\cite{pes}, if under a given transformation
the Lagrange density only changes by a total derivative,
\begin{displaymath}
  \quad D\mathcal{L}(x)\equiv 
  \frac{d\mathcal{L}(x,\lambda)}{d\lambda}\bigg|_{\lambda=0}\biggr.
  =\partial_\mu F^\mu(x),
\end{displaymath}
there is a conserved current associated with this symmetry: 
\begin{equation}
  \label{e:current}
  J^\mu(x)=\text{const.}\cdot\left\{
    \frac{\delta L_{\text s}}{\delta(\partial_\mu\phi(x))} D\phi(x)-F^\mu(x),
  \right\}
\end{equation}
where
\begin{displaymath}
  D\phi(x)\equiv\frac{d\phi(x,\lambda)}{d\lambda}\bigg|_{\lambda=0}\biggr.
\end{displaymath}
Current conservation means $\partial_\mu J^\mu=0$.  Since
(\ref{e:rot}) yields $F^\mu(x)=0$ and $D\phi(x)=1$, the particle
four-current $(J^\mu)=(c\rho,\bs{J})$ is given by
\begin{equation}
  \label{e:currents}
  J^\mu(x)
  =-\frac{1}{\hbar}\frac{\delta L_{\text s}}{\delta(\partial_\mu\phi(x))} 
  =-\frac{\delta L_{\text s}}{\delta(D_\mu\phi(x))}, 
\end{equation}
where we have chosen the normalization such that the electric current
equals the charge times the particle current:
\begin{displaymath}
  -c\;\!\frac{\delta L_{\text s}}{\delta(A_\mu(x))}
  =-e^*J^\mu(x). 
\end{displaymath}
The Lagrange density (\ref{e:ls}) yields for the particle density 
\begin{eqnarray}
  \label{e:rhocur}
  \rho(x)\!&\!=\!&\! 
  -\frac{1}{\hbar}\frac{\delta L_{\text s}}{\delta(\partial_t\phi(x))}
  \\ \label{e:rho}\rule{0pt}{16pt} 
  \!&\!=\!&\! 
  n_{\text s}-\frac{n_{\text s}}{m^*}\frac{1}{v^2}
  \Bigl(\hbar\partial_t \phi(x)-e^*\Phi(x)\Bigr)
\end{eqnarray}
% \begin{equation}
%   \label{e:rho}
%   \rho(x)
%   =-\frac{1}{\hbar}\frac{\delta L_{\text s}}{\delta(\partial_t\phi(x))}
%   =n_{\text s}-\frac{n_{\text s}}{m^*}\frac{1}{v^2}
%   \Bigl(\hbar\partial_t \phi(x)-e^*\Phi(x)\Bigr)
% \end{equation}
and for the particle current
\begin{eqnarray}
  \label{e:jcur}
  \bs{J}(x)\!&\!=\!&\!
  -\frac{1}{\hbar}\frac{\delta L_{\text s}}{\delta(\nabla\phi(x))}
  \\ \label{e:j}\rule{0pt}{16pt} 
  \!&\!=\!&\! 
  \frac{n_{\text s}}{m^*}
  \Bigl(\hbar\nabla\phi(x)+\frac{e^*}{c}\bs{A}(x)\Bigr).
\end{eqnarray}
% \begin{equation}
%   \label{e:j}
%   \bs{J}(x)
%   =-\frac{1}{\hbar}\frac{\delta L_{\text s}}{\delta(\nabla\phi(x))}
%   =\frac{n_{\text s}}{m^*}
%   \Bigl(\hbar\nabla\phi(x)+\frac{e^*}{c}\bs{A}(x)\Bigr).
% \end{equation}
The corresponding conservation law is just the continuity equation
\begin{equation}
\label{e:cont}
\partial_t\rho+\nabla\bs{J}=0.
\end{equation}

Note that since our Lagrangian (\ref{e:ltot}) does not depend on $\phi(x)$,
but only on derivatives of $\phi(x)$, \ie
\begin{displaymath}
  \frac{\delta L_{\text s}}{\delta(\phi(x))}=0,
\end{displaymath}
(\ref{e:currents}) implies that the current conservation law
\begin{displaymath}
  -\hbar\partial_\mu J^\mu(x)=\partial_\mu 
  \frac{\delta L_{\text s}}{\delta(\partial_\mu\phi(x))}=0
\end{displaymath}
is equivalent the Euler--Lagrange equation for the field $\phi(x)$.
For a neutral superfluid, we obtain 
\begin{equation}
  \left(\frac{1}{v^2}\partial_t^2 -\nabla^2\right)\phi(x)\!=0
\end{equation}
and hence the dispersion (\ref{e:lindis}) by Fourier transformation.

The most important implication of (\ref{e:currents}) for the
particle four-current is, however, that the density  
%$-\hbar c \rho(x)$ is equal to
$\rho(x)$ is up to a numerical factor equal to
the momentum field $\pi(x)$ conjugate to $\phi(x)$:
\begin{equation}
  \label{e:mom}
  -\hbar\rho(x)=\pi(x)\equiv
  \frac{\delta L_{\text s}}{\delta(\partial_t\phi(x))}. 
\end{equation}
We may hence go over to an Hamiltonian formulation, and write
the Hamiltonian density
\begin{equation}
  \label{e:h}
  \mathcal{H}_{\text s}(x)\equiv 
  -\hbar\rho(x)\:\! \partial_t\phi(x)-\mathcal{L}_{\text s}(x),
\end{equation}
which is now considered a functional of 
$\rho(x)$, $\partial_i\phi(x)$, $\Phi(x)$, %$A_0(x)$, 
and $A_i(x)$, but not
$\partial_t\phi(x)$.  (In principle, $\mathcal{H}_{\text s}$ could
also depend through $\mathcal{L}_{\text s}$ on $\phi(x)$ and $x$.
Note also that (\ref{e:h}) as the generator of time translations is
not invariant under time dependent gauge transformations, while the
equations of motions below are invariant.)  The Hamiltonian is of
course given by
\begin{equation}
  \label{e:htot}
  H_{\text s}=\int\! d^{3\,}\!\bs{x}\;\! \mathcal{H}_{\text s}(x) .
\end{equation}
Hamilton's equations are in analogy to the familiar equations
\begin{displaymath}
  \dot q=\frac{\partial H(p,q)}{\partial p},\
  \dot p=-\frac{\partial H(p,q)}{\partial q}
\end{displaymath}
from classical mechanics given by
\begin{equation}
  \label{e:h1}
  \partial_t\phi(x)=\frac{\delta H_{\text s}}{\delta (\pi(x))}
  =-\frac{1}{\hbar}\frac{\delta H_{\text s}}{\delta (\rho(x))}
\end{equation}
and
\begin{equation}
  \label{e:h2}
  -\hbar\:\!\partial_t \rho(x)=\partial_t \pi(x)=\partial_i 
  \frac{\delta H_{\text s}}{\delta (\partial_i\phi(x))}
  -\frac{\delta H_{\text s}}{\delta (\phi(x))}.
\end{equation}
With regard to the explicit equations of motion for the fields, these
equations are equivalent to the Euler--Lagrange equation.  They
provide, however, additional information regarding the physical
interpretation.  To extract this information, it is best
to study first the other conservation laws corresponding to
energy and momentum.

The theory is invariant under spacetime translations $x\rightarrow
x-e\lambda$, where $e$ is an arbitrary unit vector in spacetime (\eg
$e=(1,0,0,0)$ or $e^\nu=\delta_{0\nu}$ for a translation in time).
The infinitesimal translations are equivalent to the field and density
transformations~\cite{pes}
\begin{eqnarray}
\label{e:transl}
\phi(x)\!&\!\rightarrow\!&\!\phi(x+e\lambda)= 
\phi(x)+\lambda e^\nu\partial_\nu\phi(x),
\nonumber\\ \nonumber \rule{0pt}{15pt}
A_\mu(x)\!&\!\rightarrow\!&\! A_\mu(x+e\lambda)= 
A_\mu(x)+\lambda e^\nu\partial_\nu A_\mu(x),
\\ \nonumber \rule{0pt}{15pt}
\mathcal{L}(x)\!&\!\rightarrow\!&\!\mathcal{L}(x+e\lambda)
=\mathcal{L}(x)+\lambda\partial_\nu(e^\nu\mathcal{L}(x)),
\end{eqnarray}
which implies $D\phi=e^\nu\partial_\nu\phi$, 
$DA_\mu=e^\nu\partial_\nu A_\mu$, and $F^\mu(x)=e^\mu\mathcal{L}(x)$.
The conserved current associated  
with this symmetry is according to (\ref{e:current}) given by
\begin{displaymath}
J^\mu
= \frac{\delta L}{\delta(\partial_\mu\phi)} 
e^\nu\partial_\nu\phi
+\frac{\delta L}{\delta(\partial_\mu A_\kappa)} e^\nu\partial_\nu A_\kappa
-e^\mu\mathcal{L}
= e_\nu T_{\scriptscriptstyle\text{can}}^{\mu\nu},
\end{displaymath}
where the canonical energy-momentum tensor
$T_{\scriptscriptstyle\text{can}}^{\mu\nu}$ is the sum of the
contributions from the superfluid, the uniformly charged background,
and the electromagnetic field:
\begin{displaymath}
T_{\scriptscriptstyle\text{can}}^{\mu\nu}=
T_{\text{s},\scriptscriptstyle\text{can}}^{\mu\nu}+
T_{\text{b},\scriptscriptstyle\text{can}}^{\mu\nu}+
T_{\text{em},\scriptscriptstyle\text{can}}^{\mu\nu}
\end{displaymath}
where 
\begin{eqnarray}
\label{e:tcans}\rule{0pt}{15pt}
T_{\text{s},\scriptscriptstyle\text{can}}^{\mu\nu}\!&\!=\!&\!
\frac{\delta L_{\text s}}{\delta(\partial_\mu\phi)} 
\partial^{\:\!\nu\!}\phi
-g^{\mu\nu}\mathcal{L}_{\text s},
\\ \nonumber\rule{0pt}{16pt}
T_{\text{b},\scriptscriptstyle\text{can}}^{\mu\nu}\!&\!=\!&\!
-g^{\mu\nu}\mathcal{L}_{\text b},
\\ \label{e:tcane}\rule{0pt}{20pt}
T_{\text{em},\scriptscriptstyle\text{can}}^{\mu\nu}\!&\!=\!&\!
\frac{\delta L_{\text{em}}}{\delta(\partial_\mu A_\kappa)} 
\partial^{\:\!\nu\!}A_\kappa
-g^{\mu\nu}\mathcal{L}_{\text{em}}.
\end{eqnarray}
% Only the total current is conserved,
% $\partial_\mu T_{\scriptscriptstyle\text{can}}^{\mu\nu}=0$. 
The conservation law $\partial_\mu
T_{\scriptscriptstyle\text{can}}^{\mu\nu}=0$ describes energy
conservation for $\nu=0$ and momentum conservation for $\nu=i$.  The
$\mu=0$ components of $T_{\scriptscriptstyle\text{can}}^{\mu\nu}$
correspond to energy and momentum densities; in particular, $-c\int\!
d^{3\,}\!x\, T_{\scriptscriptstyle\text{can}}^{00}$ generates
translations in time and $\int\! d^{3\,}\!x\,
T_{\scriptscriptstyle\text{can}}^{0i}$ translations in space.

In the case of a gauge theory, like the theory of a charged superfluid
we consider here, it is not possible to interpret
$T_{\text{s,}\scriptscriptstyle\text{can}}^{00}$ as the energy or
$\frac{1}{c}T_{\text{s,}\scriptscriptstyle\text{can}}^{0i}$ as the
kinematical momentum density of the superfluid.  The reason is simply
that (\ref{e:tcans}) (and also (\ref{e:tcane})) is not gauge
invariant.  To circumvent this problem, we simply supplement the naive
translations by a suitable gauge transformation, such that the fields
transform covariantly:
\begin{eqnarray}
\phi&\!\rightarrow\!&\! \phi+\lambda 
e^\nu\Bigl(\partial_\nu\phi-\frac{e^*}{\hbar c}A_\nu\Bigr),
\nonumber \\ \nonumber \rule{0pt}{15pt}
A_\kappa\!\!&\!\rightarrow\!&\! A_\kappa+\lambda 
e^\nu (\partial_\nu A_\kappa-\partial_\kappa A_\nu).
\end{eqnarray}
The gauge transformation is hence given by (\ref{e:gauge}) with
$\Lambda(x)=\lambda e^\nu A_\nu(x)$.  This yields the ``kinematical''
energy momentum tensor, with contributions from the superfluid and the
electromagnetic field%~\cite{gww}
\begin{eqnarray}
  \label{e:tensor}
  T_{\text s}^{\mu\nu}\!&\!=\!&\!
  \frac{\delta L_{\text s}}{\delta(D_\mu\phi)} D^{\:\!\nu\!}\phi
  -g^{\mu\nu}_{\phantom{\mu\nu}}\mathcal{L}_{\text s},
  \label{e:ts} \\ \nonumber \rule{0pt}{20pt}
  T_{\text{em}}^{\mu\nu}\!&\!=\!&\!
  \frac{\delta L_{\text{em}}}{\delta(F_{\mu\kappa})} 
  F_{\phantom{\:\!\nu}\kappa}^{\:\!\nu}
  -g^{\mu\nu}_{\phantom{\mu\nu}}\mathcal{L}_{\text{em}}.
\end{eqnarray}
These expressions are manifestly gauge invariant.  Using
(\ref{e:currents}) with $\mu=0$ or (\ref{e:rhocur}), 
we can write the energy density of the superfluid
\begin{equation}
  \label{e:eden}
  T_{\text s}^{00}(x)= -\rho(x)
  \Bigl(\hbar\partial_t\phi(x) -e^* \Phi(x)\Bigr)-\mathcal{L}_{\text s}(x).
\end{equation}
Note that this is numerically equal to
\begin{equation}
  \label{e:eden1}
  T_{\text s}^{00}(x)=\mathcal{H}_{\text s}(x)+\rho(x)e^* \Phi(x).
\end{equation}
Similarly, we can write the momentum density
\begin{equation}
\label{e:mden}
\frac{1}{c}T_{\text s}^{0i}(x)=
\rho(x)\Bigl(\hbar\nabla\phi(x)+\frac{e^*}{c}\bs{A}(x)\Bigr).
\end{equation}

We can use this expression to introduce the superfluid velocity
$\bs{v}_{\text s}(x)$.  In terms of $\bs{v}_{\text s}(x)$, the momentum
density of the superfluid has to be given by
\begin{equation}
  \label{e:tvs}
  \frac{1}{c}T_{\text s}^{0i}(x)\stackrel{!}{=}
  \rho(x)\, m^*\;\!\!\bs{v}_{\text s}(x),
\end{equation}
which leads us to define
\begin{equation}
  \label{e:vs}
  m^*\bs{v}_{\text s}(x)\equiv
  \hbar\nabla\phi(x)+\frac{e^*}{c}\bs{A}(x)=D_i\phi(x).
\end{equation}
Since $\bs{v}_{\text s}(x)$ is to be interpreted as a physical velocity, it
has to transform like a velocity under a Galilean transformation, 
\begin{equation}
  \label{e:gal1}
  \bs{v}_{\text s}(x)\ \rightarrow\ \bs{v}_{\text s}(x) + \bs{u}.
\end{equation}
The total momentum of the superfluid will hence transform according to
\begin{equation}
  \label{e:gal2}
  \frac{1}{c}\int\! d^{3\,}\!\bs{x}\, T_{\text s}^{0i}(x)
  \ \rightarrow\ \frac{1}{c}\int\! d^{3\,}\!\bs{x}\, T_{\text s}^{0i}(x) 
  +\!\int\! d^{3\,}\!\bs{x}\:\!\rho(x)\  m^*\bs{u},
\end{equation}
which implies directly that in a translationally invariant system,
$m^*$ has to be the bare mass of the superfluid particles or Cooper
pairs~\cite{conn}.

It should also be possible to express the particle current in terms 
of the superfluid velocity.  
Since the same particles which carry the momentum also
carry the current, the particle current has to be given by 
\begin{equation}
\label{e:js}
\bs{J}(x)=\rho(x)\:\!\bs{v}_{\text s}(x).
\end{equation}
This is almost, but not quite, equivalent to our earlier expression
(\ref{e:j}), as $n_{\text s}$ is only the leading contribution to
$\rho(x)$.  So either (\ref{e:js}) with (\ref{e:vs}) or (\ref{e:j}) is
not fully correct.  To see which one, recall that we have only used
the general expression for the density %(\ref{e:currents}) with $\mu=0$
(\ref{e:rhocur})
as defined through particle number conservation %(\ref{e:currents}) 
in obtaining (\ref{e:mden}) and hence (\ref{e:vs}) and (\ref{e:js})
from (\ref{e:ts}), while we have used the explicit expression for the
Lagrange density (\ref{e:ls}) in obtaining (\ref{e:j}).  In other
words, only symmetry considerations enter in (\ref{e:mden}), while
(\ref{e:j}) depends explicitly on the Lagrange density.  The
expression for the momentum density (\ref{e:mden}), and hence our
definition of the superfluid velocity (\ref{e:vs}), is therefore
exact, while the expression (\ref{e:j}) for the particle current is
only an approximation~\cite{phila}.

The expression for the current, however, will assume the exact and
physically correct form (\ref{e:js}) if we introduce suitable
corrections to the effective Lagrangian.  To obtain these, we simply
require the Lagrangian to satisfy~\cite{gww}
\begin{equation}
  \label{e:gww}
  \frac{1}{c}\:\!T_{\text s}^{0i}(x)= m^*J^i(x)
\end{equation}
or
\begin{equation}
\frac{1}{c}\frac{\delta L_{\text s}}{\delta(D_0\phi)} D^{\:\!i\!}\phi
 = -m^*\frac{\delta L_{\text s}}{\delta(D_i\phi)}.
 \end{equation}
Upon integration of this equation we find that the Lagrange density
must be of the form
\begin{equation}
\mathcal{L}_{\text s}
=P\Bigl(cD_0\phi+\frac{1}{2m^*} (D_i\phi)^2\Bigr),
\end{equation}
where $P$ is an arbitrary polynomial.
Our superfluid Lagrange density (\ref{e:ls2}) will assume this form if we
add third and fourth order corrections in $D_\mu\phi$; the full
superfluid Lagrange density is then given by~\cite{gww,schakel}
\begin{eqnarray}
\label{e:ls3}
\mathcal{L}_{\text s}\!&\!=\!&\!-\,n_{\text{s}}\;\!
\Bigl(cD_0\phi+\frac{1}{2m^*} (D_i\phi)^2\Bigr)\cr\rule{0pt}{18pt}
&&\!+\,\frac{n_{\text{s}}}{2m^*}\frac{1}{v^2}\;\!
\Bigl(cD_0\phi+\frac{1}{2m^*} (D_i\phi)^2\Bigr)^2.
\end{eqnarray}
It yields for the particle density 
\begin{eqnarray}
\label{e:rho3}
\rho(x)\!&\!=\!&\!
-\frac{1}{c}\frac{\delta L_{\text s}}{\delta(D_0\phi)}
\nonumber \\ \rule{0pt}{17pt}
%\!&\!=\!&\!
&& \hspace{-10mm}=
n_{\text s}-\frac{n_{\text s}}{m^*}\frac{1}{v^2}
\Bigl(cD_0\phi+\frac{1}{2m^*}(D_i\phi)^2\Bigr)
\\ \nonumber \rule{0pt}{17pt}
%\!&\!=\!&\!
&& \hspace{-10mm}=
n_{\text s}-\frac{n_{\text s}}{m^*}\frac{1}{v^2}
\biggl\{\Bigl(\hbar\partial_t \phi-e^*\Phi\Bigr)+\frac{1}{2m^*}
\Bigl(\hbar\nabla\phi+\frac{e^*}{c}\bs{A}\Bigl)^2\biggr\}
\end{eqnarray}
and for the particle current
\begin{equation}
\label{e:j3}
\bs{J}(x)=-\frac{\delta L_{\text s}}{\delta(D_i\phi)}
=\rho(x)\:\!\bs{v}_{\text s}(x),
\end{equation}
where $\rho(x)$ and $\bs{v}_{\text s}(x)$ are given by (\ref{e:rho3}) and
(\ref{e:vs}).

Let us now return to Hamilton's equations, and in particular their
physical interpretation.  With (\ref{e:eden1}) we may rewrite
(\ref{e:h1}) as
\begin{eqnarray}
\label{e:h1'}
\hbar\partial_t\phi(x)
\!&\!=\!&\!-\frac{\delta H_{\text s}}{\delta (\rho(x))}
=-\frac{\partial\mathcal{H}_{\text s}(x)}{\partial\rho(x)}
=-\frac{\partial T_{\text s}^{00}(x)}{\partial\rho(x)}+e^*\Phi(x)
\nonumber\\ %\nonumber
\rule{0pt}{14pt}
\!&\!\!=\!\!&\!\!-\mu(x)+e^*\Phi(x),
\end{eqnarray}
where we have used the definition of the chemical potential.  This is one
of two equations Anderson~\cite{phil66} refers to as ``characteristic of
superfluidity''.  In analogy to the definition (\ref{e:vs}) of
$\bs{v}_{\text s}(x)$, we rewrite it for later purposes as
\begin{equation}
\label{e:mu}
-\mu(x)=\hbar\partial_t\phi(x)-e^*\Phi(x)=cD_0\phi(x).
\end{equation}
Taking the gradient and adding $\frac{e^*}{c}\partial_t\bs{A}(x)$ 
on both sides of (\ref{e:h1'}), we obtain 
\begin{eqnarray}
  \label{e:muelch}
  \partial_t\Bigl(\hbar\nabla\phi(x)+\frac{e^*}{c}\bs{A}(x)\Bigr)
  \!&\!=\!&\!-\nabla\mu(x)-e^*\bs{E}(x)\cr \rule{0pt}{16pt}
  \!&\!=\!&\!-\nabla\mu_{\scriptscriptstyle\text{el.chem.}}(x),
\end{eqnarray}
where we used the definitions of the electric field,
\begin{displaymath}
\bs{E}\equiv -\nabla\Phi-\frac{1}{c}\partial_t\bs{A},
\end{displaymath}
and of the electrochemical potential.  With (\ref{e:vs}) we may write  
\begin{equation}
\label{e:acc}
m^*\:\!\partial_t\bs{v}_{\text s}(x)
=-\nabla\mu_{\scriptscriptstyle\text{el.chem.}}(x).
\end{equation}
The gradient of the electrochemical potential (or chemical potential
for a neutral superfluid) is usually~\cite{phil66} identified with
minus the total force on the particles, and (\ref{e:acc}) is referred
to as the ``acceleration equation''.  This is, however, not quite
correct.  $\partial_t \bs{v}_{\text s}$ in (\ref{e:acc}) denotes the
time derivative in the superfluid velocity field at spacetime $x$
(known as ``local acceleration'' in hydrodynamics), while the force on
the particles is given by the time derivative of the velocity of a
given particle in the fluid at $x$ (``substantial acceleration'' in
hydrodynamics):
\begin{equation}
\label{e:force}
\frac{1}{m^*}\bs{F}=\frac{d\bs{v}_{\text s}}{dt}
=\partial_t\bs{v}_{\text s}+(\bs{v}_{\text s}\nabla)\bs{v}_{\text s}.
\end{equation}
Nonetheless, (\ref{e:acc}) is one of the fundamental equations in the
phenomenology of superfluidity.  It states %, for one thing, 
that if
there is a gradient in the electrochemical (or chemical) %for $e^*=0$) 
potential in a superconductor (or superfluid), the superfluid
will be ``accelerated'' without any frictional damping.  On the other
hand, if the superfluid flow is stationary, the electrochemical (or
chemical) potential has to be constant across the superconductor (or
superfluid).  Since a voltmeter measures a difference in the
electrochemical potential, there cannot be a voltage across a
superconductor unless the flow is ``accelerated''.

Let us now turn to Hamilton's second equation (\ref{e:h2}).  We first
rewrite it as
\begin{equation}
\label{e:h2'}
-\hbar\:\!\partial_t\rho(x)
=\partial_i \frac{\delta H_{\text s}}{\delta (\partial_i\phi(x))}
-\frac{\partial\mathcal{H}_{\text s}(x)}{\partial\phi(x)}.
\end{equation}
Since the last term in (\ref{e:eden1}), $\rho(x)e^* \Phi(x)$, does not
depend on the phase $\phi(x)$, we may replace $\mathcal{H}_{\text
  s}(x)$ in the last term in (\ref{e:h2'}) by $T_{\text s}^{00}(x)$.
Integrating the resulting equation over the superfluid, discarding a
boundary term, and defining a ``global'' derivative with respect to
the phase,
\begin{displaymath}
  \frac{\partial F[\phi(x)]}{\partial\phi}\equiv
\lim_{\Delta\phi\rightarrow 0}\frac{F[\phi(x)+\Delta\phi]}{\Delta\phi},
\end{displaymath}
where $F[\phi(x)]$ is an arbitrary functional of $\phi(x)$
and $\Delta\phi$ an infinitesimal independent of spacetime, yields
\begin{displaymath}
-\hbar\:\!\partial_t \int\!d^{3\,}\!{\bs x}\,\rho(x)
=-\frac{\partial}{\partial\phi}\,\int\!\! 
d^{3\,}\!{\bs x}\:T_{\text s}^{00}(x)\,
\end{displaymath}
or
\begin{equation}
\label{e:h2''}
\hbar\:\!\partial_t N=\frac{\partial E_{\text s}}{\partial\phi}\:\!,
\end{equation}
where $N$ is the number of particles (or pairs) and $E_{\text s}$ the
energy of the superfluid.  This is the other ``characteristic
equation'' of superfluidity~\cite{phil66}.

This concludes our derivation or motivation of the fundamental
equations of superfluidity in the limit of low temperatures and low
energies.  Before turning to the phenomenology these equations imply,
I would like to digress briefly and justify one of the implicit
assumptions made above.  The assumption is that we can describe the
macroscopic quantum phenomena of superfluidity with a classical
effective field theory, or in other words, that we may consider both
the phase $\phi(x)$ and its conjugate field $\pi(x)=-\hbar\rho(x)$ as
thermodynamic variables.  To justify this assumption, let us
canonically quantize the theory by imposing
\begin{equation}
  \label{e:canq}
  \bigc{\hat\phi(\bs{x},t)}{\hat\pi(\bs{y},t)}=i\hbar\delta(\bs{x}-\bs{y}).
\end{equation}
Integration of $\bs{y}$ over the superfluid yields
\begin{displaymath}
  \label{e:canq2}
  \bigc{\hat\phi(\bs{x})}{\hat N}=-i,
\end{displaymath}
which in turn implies the uncertainty relation
\begin{displaymath}
  \Delta\phi(\bs{x})\;\!\Delta N\ge\,\frac{1}{2}.
\end{displaymath}
If we assume that the number of particles in the superfluid takes on a
macroscopic value of order $N\approx 10^{20}$, a $\Delta N$ of the
order of $\sqrt{N}$ implies a relative uncertainty of order $10^{-10}$
in the particle number and the phase.  These numbers are comparable to
the position and momentum uncertainties of a macroscopic object.  The
description of a macroscopic superfluid in terms of a classical field
theory is therefore as appropriate as the classical description of any
other macroscopic object.  This is of course not in contradiction with
the fact that Planck's constant $\hbar$ appears in this effective
field theory.  We will see below that it manifests itself in a family
of ``quantum effects'', which are related to the compactness of the
U(1) field $\phi(x)$.  These effects require either a non-trivial
topology or more than one superfluid, and are very similar for neutral
and for charged superfluids.

%\section{PHENOMENOLOGY OF SUPERFLUIDITY AND THE HIGGS MECHANISM}
%\section{Phenomenology of superfluidity and the Higgs mechanism}
\section{PHENOMENOLOGY AND THE HIGGS MECHANISM}
%\section{Phenomenology and the Higgs mechanism}
\label{sec:phe}

To begin with, however, let us consider the superfluid flow in a
simply connected superfluid.  The phenomenology depends strikingly on
whether the fluid is charged or not.  For a neutral superfluid,
$e^*=0$, and the gauge field decouples completely.  Even for a fixed
set of boundary conditions, we have an infinite set of solutions for
the superfluid flow, corresponding via
\begin{displaymath}
m^*\bs{v}_{\text s}(x)=\hbar\nabla\phi(x)  
\end{displaymath}
to all possible choices of the phase field $\phi(x)$.  In a simply
connected superfluid, the flow will be vortex-free, \ie
\begin{displaymath}
  \nabla\times\bs{v}_{\text s}(x)=0,
\end{displaymath}
and subject to boundary constraints, but apart from this, it only has
to satisfy the continuity equation as an equation of motion.

The simplest example of a multiply connected superfluid is a
superfluid with a line defect, or vortex, along which the magnitude
$|\Psi(x)|$ of the superfluid order parameter vanishes.  The
phase $\phi(x)$ still has to be single valued everywhere in the fluid,
but being a phase, its value may change by a multiple of $2\pi$ as we
circumvent the line defect along a closed curve $\partial S$:
\begin{equation}
  \label{e:vortex}
  \oint_{\partial S}\!\nabla\phi(x)d\bs{l}=2\pi n
\end{equation}
where $n$ is an integer.  The angular momentum of each superfluid
``particle'' around the vortex is hence quantized in units of
$\hbar$.  With Stokes theorem and the definition
\begin{equation}
  \label{e:vorti}
  \bs{\omega}(x)=\nabla\times\bs{v}_{\text s}(x),
\end{equation}
we may express this alternatively as quantization condition for the 
vorticity
\begin{equation}
  \label{e:vortiq}
  \int_{S}\bs{\omega}(x)\!\cdot\!\bs{n} da =\frac{2\pi\hbar\;\!n}{2m^*},
\end{equation}
where $\bs{n}$ is a unit vector normal to the surface and the area
integral extends over any open surface $S$ which is pierced by the
vortex once.  The quantization of vortices in a superfluid is the
simplest of the ``quantum effects'' alluded to above, where Planck's
constant $\hbar$ enters in the phenomenology through the compactness
of the field $\phi(x)$.  (If $\phi(x)$ was not compact, we could
eliminate $\hbar$ completely from the effective theory by rescaling
$\phi(x)\rightarrow \hbar\phi(x)$.)

Let us now turn to the phenomenology of a simply connected charged
superfluid or superconductor, which displays the Higgs mechanism.  The
essence of the mechanism is that the phase field $\phi(x)$ looses its
independent significance in the presence of the gauge field.  
There are two ways of seeing this.  The first is on the level of the
equations of motion.  We can simply choose a gauge such that
$\phi(x)=0$ everywhere in the fluid; any other choice of gauge can be
brought into this gauge via %a gauge transformation 
(\ref{e:gauge}) with 
\begin{displaymath}
  \Lambda(x)=\frac{\hbar c}{e^*}\phi(x).
\end{displaymath}
The second is on the level of the effective Lagrangian.  We may
introduce a new vector field
\begin{equation}
  \label{e:newfields}
  -\frac{e^*}{c}A_\mu'\equiv
  D_\mu\phi=\hbar\:\! \partial_\mu \phi-\frac{e^*}{c}A_\mu\:\!. 
\end{equation}
In terms of this field, the superfluid Lagrange density (\ref{e:ls3})
looks the same except that all terms containing $\phi$ have
disappeared.  In particular, the terms quadratic in the derivatives of
$\phi$ in $\mathcal{L}_{\text s}$ have turned into a mass term
\begin{equation}
  \label{e:mass}
  \frac{n_{\text{s}}}{2m^*}\biggl\{\frac{1}{v^2}
  \left(e^*A_0'\right)^2-
  \Big(\frac{e^*}{c}\bs{A}'\Bigl)^2\biggr\}
\end{equation}
for the vector field.  The Maxwell Lagrange density and the Lagrange
density for the uniform neutralizing background charge take the same
form with $F_{\mu\nu}$ and $A_\mu$ replaced by $F_{\mu\nu}'$ and
$A_\mu'$, respectively, except for a total derivative or boundary term
we discard.  Thus the massless gauge field $A_\mu$ is replaced by a
massive vector field $A_\mu'$, while the Goldstone boson $\phi$
disappeared.  The total number of degrees of freedom, however, is
preserved: before, the massless vector field has two (the two helicity
states of the photon) and the Goldstone boson one, while the massive
vector field after the change of variables has three degrees of
freedom.  In Sidney Coleman's words, ``the vector field has eaten the
Goldstone bosons and grown heavy''~\cite{sid}.  We will return to this
issue after studying the phenomenology of the superconductor using the
equations of motion.

The Euler--Lagrange equation for $A_\mu$,
\begin{equation}
  \label{e:eulag}\nonumber
  \partial_\mu \frac{\delta L}{\delta(\partial_\mu A_\nu)}
  -\frac{\delta L}{\delta A_\nu}=0,
\end{equation}
yields Maxwell's electrodynamics with electric charge density
$-e^*(\rho-n_{\text s})$ and current $-e^*\bs{J}$,
\begin{eqnarray}
  \nabla\cdot\bs{E}\!&\!=\!&\!-4\pi e^*(\rho-n_{\text s}),
  \label{e:max1} \\ \rule{0pt}{20pt} \label{e:max2}
  \nabla\times\bs{B}-\frac{1}{c}\partial_t\bs{E}
  \!&\!=\!&\!-\frac{4\pi e^*}{c} \bs{J},
\end{eqnarray}
where 
\begin{equation}
  \label{e:eandb}\nonumber
  \bs{E}=-\nabla\Phi-\frac{1}{c}\partial_t\bs{A},\
  \bs{B}=\nabla\times\bs{A},
\end{equation}
and $\rho$ and $\bs{J}$ are given by (\ref{e:rho3}) and (\ref{e:j3}),
respectively.  In principle, we could also obtain the continuity
equation (\ref{e:cont}) as the Euler--Lagrange equation for $\phi$,
but since $L$ depends on derivatives of $A_\mu$ only through
$F_{\mu\nu}$ and $\phi$ is minimally coupled to $A_\mu$,
(\ref{e:cont}) is automatically satisfied by any solution of
(\ref{e:max1}) and (\ref{e:max2}).  This is consistent with the fact
that $\phi$ has lost its independent significance due to the Higgs
mechanism.

For convenience, we choose the gauge $\phi(x)=0$.  Then (\ref{e:rho3}), 
(\ref{e:j3}), and (\ref{e:vs}) imply
\begin{equation}
  4\pi e^*(\rho-n_{\text s})=
  \frac{4\pi{e^*}^2 n_{\text s}}{m^*v^2}
  \Bigl(\Phi-\frac{e^*}{2m^*c^2}\bs{A}^2\Bigl),
  \label{e:rho4} 
\end{equation}
\vspace{-2pt}
\begin{equation}
  \label{e:j4}
  \frac{4\pi e^*}{c}\bs{J}  =
  \frac{4\pi{e^*}^2 n_{\text s}}{m^*c^2}\bs{A}
  \biggl\{1+\frac{e^*}{m^*v^2}
  \Bigl(\Phi-\frac{e^*}{2m^*c^2}\bs{A}^2\Bigl)\biggr\}.
\end{equation}
Let us now restrict our attention to quasistatic phenomena, where we
can neglect the time derivative terms.  The analysis given below implies
that this assumption holds for frequencies significantly smaller than
$c/\lambda_{\text L}$, where 
\begin{equation}
\lambda_{\text L} = \sqrt{\frac{m^*c^2}{4\pi{e^*}^2 n_{\text s}}} 
%\frac{1}{\lambda_{\text L}^2}=\frac{4\pi{e^*}^2 n_{\text s}}{m^*c^2} 
\end{equation}
is the London penetration depth.
Then %the equations 
(\ref{e:max1})--(\ref{e:j4}) reduce to
\begin{eqnarray}
  \nabla^2\Phi \!&\!=\!&\! 
  \frac{c^2}{v^2}\frac{1}{\lambda_{\text L}^{\;2}}
  \Bigl(\Phi-\frac{{e^*}}{2m^*c^2}\bs{A}^2\Bigl),
  \label{e:eom1} \\ \rule{0pt}{20pt} \label{e:eom2}
  \nabla^2\bs{A}\!-\!\nabla(\nabla\bs{A})\!&\!=\!&\! 
  \frac{1}{\lambda_{\text L}^{\;2}}
  \bs{A}\biggl\{1\!+\!\frac{e^*}{m^*v^2}
  \Bigl(\Phi\!-\!\frac{{e^*}}{2m^*c^2}\bs{A}^2\Bigl)\!\biggr\}.\qquad
\end{eqnarray}
Let us first look at the linear terms in these equations, \ie the
solution for infinitesimal $\Phi$ and $\bs{A}$.  Under quasistatic
conditions, (\ref{e:max2}) implies $\nabla\bs{J}=0$ and with
(\ref{e:j4}) for infinitesimal fields $\nabla\bs{A}=0$.  The equations
reduce to
\begin{eqnarray}
  \nabla^2\Phi \!&\!=\!&\! 
  \frac{c^2}{v^2}\frac{1}{\lambda_{\text L}^{\;2}}\Phi,
  \label{e:eomi1}\\ \rule{0pt}{20pt} \label{e:eomi2}
  \nabla^2\bs{A}\!&\!=\!&\! 
  \frac{1}{\lambda_{\text L}^{\;2}}
  \bs{A},
\end{eqnarray}
\ie we have electric screening in addition to magnetic screening, but
with a screening length reduced by a factor $v/c$.  This leads us to
conjecture that the dominant energy is the Coulomb interaction, which
effects charge neutrality or $\rho(x)\approx n_{\text s}$.  We now
simply assume that this is a valid approximation, and justify it {\sl
  a posteriori}.  Then (\ref{e:rho4}) implies
\begin{equation}
  \label{e:hall}
   \Phi-\frac{{e^*}}{2m^*c^2}\bs{A}^2=0,
\end{equation}
and (\ref{e:j4}) reduces to
\begin{equation}
  \label{e:j5}
  \bs{J}=\frac{e^* n_{\text s}}{m^*c}\bs{A}.
\end{equation}
Taking the curl of this equation, we obtain London's
equation~\cite{london,london1}
\begin{equation}
  \label{e:lon2}
  \nabla\times\bs{J}=\frac{e^* n_{\text s}}{m^*c}\bs{B}.
\end{equation}
Under quasistatic conditions, we have again $\nabla\bs{J}=0$ and with
(\ref{e:j5}) $\nabla\bs{A}=0$, which implies that (\ref{e:eom2})
reduces to (\ref{e:eomi2}).  The solution of (\ref{e:eomi2}) describes
exponential screening with penetration depth $\lambda_{\text L}$.  If
we have, for example, a superconductor which occupies the half-space
$x>0$ subject to an external magnetic field $\bs{B}=B_0\bs{\hat y}$ at
the boundary $x=0$, we obtain
\begin{equation}
  \label{e:sol}
  \bs{A}=A_0\;\! e^{-x/\lambda_{\text L}}\bs{\hat z},\ 
  \bs{B}=B_0\;\! e^{-x/\lambda_{\text L}}\bs{\hat y},\
%  \ \text{and}\
  \bs{J}=J_0\;\! e^{-x/\lambda_{\text L}}\bs{\hat z},
\end{equation}
where
\begin{displaymath}
  A_0=\lambda_{\text L} B_0,\    
  J_0=\sqrt{\frac{n_{\text s}}{4\pi m^*}}\;\! B_0.
\end{displaymath}
The screening of the magnetic field is known as the Meissner effect.
According to (\ref{e:hall}), the vector potential implies
an electrostatic potential 
\begin{equation}
  \label{e:hall2}
   \Phi%(x)
   =\frac{{B_0}^2}{8\pi e^* n_{\text s}}\;\! e^{-2x/\lambda_{\text L}}.
\end{equation}
This potential allows us to verify the validity of our approximation
$\rho(x)\approx n_{\text s}$.  Substituting (\ref{e:hall2}) into
(\ref{e:eomi1}), we find that the ratio of the neglected term to the
terms kept is
\begin{equation}
  \label{e:approx}
  \frac{\nabla^2\Phi}{\frac{c^2}{v^2}\frac{1}{\lambda_{\text L}^{\;2}}\Phi}
  =\frac{4 v^2}{c^2}%\approx 10^{-6}, 
  \ll 1,
\end{equation}
\ie the approximation is excellent.

The electrostatic potential (\ref{e:hall2}) is called the London or
Bernoulli Hall effect~\cite{london,gww}.  To understand its physical
origin, it is best to rewrite (\ref{e:hall}) with (\ref{e:vs}) for
$\phi(x)=0$ in terms of the superfluid velocity:
\begin{equation}
  \label{e:hall4}
  -e^*\Phi+\frac{1}{2}m^*{\bs{v}_{\text s}}^2=0.
\end{equation}
The electrostatic potential simply compensates the kinetic energy
contribution to the chemical potential, as required by (\ref{e:h1'})
with $\phi(x)=0$.  For stationary flow, this condition reduces to the
requirement that the electrochemical potential
$\mu_{\scriptscriptstyle\text{el.chem.}}$ is constant across the
superconductor.  In practice, the London Hall effect can only be
measured with capacitive contacts, as ohmic contacts are sensitive to
the electrochemical rather than the electrostatic potential~\cite{bok}.
The effect furnishes us with an independent meaning of the superfluid
density $n_{\text s}$ or the effective mass $m^*$, while under
quasistatic conditions all other effects~\cite{rot} depend only on the
superfluid stiffness $n_{\text s}/m^*$.  The underlying theoretical
reason is that the London Hall effect is a consequence of the
corrections incorporated in the effective Lagrange density
(\ref{e:ls3}).  In these terms, the parameter $m^*$ enters by itself,
while (apart from a total time derivative term irrelevant to the
equations of motion) only the combination $n_{\text s}/m^*$ entered in
the previous approximative Lagrangian (\ref{e:ltot}) with
(\ref{e:ls}), (\ref{e:lb}) and (\ref{e:lmax}).

Since we have given a precise definition of the superfluid velocity,
it is legitimate to ask whether the London Hall effect (\ref{e:hall4})
balances the Lorentz force, or, if not, what other forces balance it.
The total electromagnetic force on a given
particle with charge $-e^*$ in the fluid is given by
\begin{eqnarray}
   \bs{F}_{\text{em}}\!&\!=\!&\! 
   -e^*\bigl(\bs{E}
   +\frac{\bs{v}_{\text s}}{c}\times\bs{B}\bigr)
   \label{e:lorentz} \\ \rule{0pt}{20pt} \nonumber
   \!&\!=\!&\!
   -e^*\Bigl(-\frac{1}{c}\partial_t\bs{A}-\nabla\Phi
   +\frac{\bs{v}_{\text s}}{c}\times (\nabla\times\bs{A})\Bigr).
\end{eqnarray}
With (\ref{e:hall4}) and (\ref{e:vs}) we obtain 
\begin{eqnarray}
  \bs{F}_{\text{em}}\!&\!=\!&\! 
  m^*\Bigl(\partial_t\bs{v}_{\text s}
  +\frac{1}{2}\nabla (\bs{v}_{\text s})^2
  -\bs{v}_{\text s}\times (\nabla\times\bs{v}_{\text s})\Bigr)
  \label{e:lorentz2}\nonumber \\ \rule{0pt}{20pt} \nonumber
  \!&\!=\!&\!
  m^*\bigl(\partial_t\bs{v}_{\text s}
  +(\bs{v}_{\text s}\nabla )\bs{v}_{\text s}\bigr)
  =m^*\frac{d\bs{v}_{\text s}}{dt}.
\end{eqnarray}
Thus the gradient $\nabla\Phi$ of the electrostatic potential
(\ref{e:hall4}) does not balance the Lorentz force
$\frac{1}{c}\bs{v}_{\text s}\times\bs{B}$, but both terms together
account for the difference between local and substantial acceleration
(\ref{e:force}) in the superfluid.  This difference is significant
when, for example, the flow is stationary but does not follow a
straight line.

Let us summarize how the Higgs mechanism manifests itself in the
equations of motion.  For a neutral superfluid with $e^*=0$, there are
many solutions to
\begin{displaymath}
  \bs{J}(x)=\rho\bs{v}_{\text s}
  =\frac{\rho}{m^*}\Bigl(\hbar\nabla\phi+\frac{e^*}{c}\bs{A}\Bigr)
\end{displaymath}
for fixed boundary conditions, corresponding to all possible
configurations for the phase field $\phi(x)$.  These solutions reflect
the existence of the massless sound mode described by the field
$\phi(x)$.  For the charged superfluid, however, all the different
configurations for $\phi(x)$ merely correspond to different choices of
gauge; as far as the current or magnetic field distributions are
concerned, all these solutions are always equivalent to one for which
we have $\phi(x)=0$.  Thus the field $\phi$ does no longer describe an
excitation.  For a simply connected superconductor, it is only
meaningful in that it assures gauge invariance, both on the level of
the Lagrange density and on the level of the equations of motion.  The
solution of these equations is physically (\ie apart from the freedom
to choose the gauge) unique for a given set of boundary conditions.

We now return to the manifestation of the Higgs mechanism on the level
of the Lagrangian.  For a simply connected superconductor, we have
already seen that we may eliminate the phase field $\phi$ if we
introduce a new vector field $A_\mu'$ according to
(\ref{e:newfields}).  The mass term (\ref{e:mass}) we find for
$A_\mu'$ may appear to violate gauge invariance, as mass terms
generally do, and may, at first sight, to be taken as a signature of a
spontaneously broken gauge invariance.  For one thing, however, gauge
invariance is not violated.  From the definition (\ref{e:newfields})
it is clear that the new field simply transforms as
\begin{displaymath}
  A_\mu'(x)\rightarrow A_\mu'(x)
\end{displaymath}
under a gauge transformation (\ref{e:gauge}).  The Lagrangian hence
remains manifestly gauge invariant, and has to remain gauge invariant,
as it is the same Lagrangian as before expressed in terms of different
fields.  Furthermore, if a symmetry is spontaneously broken, it is never
violated on the level of the Lagrangian or the Hamiltonian, but only
on the level of the ground state.

In the literature, one sometimes finds the statement that ``the gauge
field acquires a mass'' due to the Higgs mechanism.  This is not
exactly to the point, as it suggests that the massive vector field
$A_\mu'$ is still a gauge field, while we have just seen that it is
gauge invariant.  In the case of a superconductor, we even know how to
interpret the individual components of $A_\mu'$ physically.
According to (\ref{e:newfields}), (\ref{e:mu}), and (\ref{e:vs}),
\begin{equation}
  \label{eq:a'}
  -\frac{e^*}{c}\left(A_\mu'\right)
  =\bigl(-\frac{\mu}{c},m^*\bs{v}_{\text s}\bigr).
\end{equation}
The Higgs mechanism hence does not imply that ``the electromagnetic
gauge field acquires a mass'', but only that we can describe the
superconductor in terms of gauge invariant fields, that is, in
terms of the chemical potential $\mu(x)$ and the superfluid velocity
$\bs{v}_{\text s}(x)$.  If we do this, we also have to express the
Maxwell Lagrange density (\ref{e:lmax}) in terms of $\mu$ and
$\bs{v}_{\text s}$.  With
\begin{displaymath}
  -\frac{1}{2}F_{\mu\nu}F^{\mu\nu}=\bs{E}^2-\bs{B}^2
\end{displaymath}
we obtain for the total Lagrange density 
\begin{eqnarray}
  \mathcal{L}%_{\text{tot}}
  \!&\!=\!&\!
  \frac{1}{8\pi {e^*}^2}
  \Bigl\{(\nabla\mu+m^*\partial_t\bs{v}_{\text s})^2-
  c^2{m^*}^2 (\nabla\times\bs{v}_{\text s})^2\Bigr\}
  \nonumber \\ \nonumber
   \rule{0pt}{20pt}
  &&\hspace{-7mm} -n_{\text{s}}\biggl\{
  \mu+\Bigl(-\mu+\frac{1}{2}m^*\bs{v}_{\text s}^2\Bigr)
  +\frac{1}{2m^*}\frac{1}{v^2}
  \Bigl(-\mu+\frac{1}{2}m^*\bs{v}_{\text s}^2\Bigr)^2\biggr\}.\\
  \label{e:l2}
\end{eqnarray}
The Euler--Lagrange equations we obtain from (\ref{e:l2}) for $\mu$
and $\bs{v}_{\text s}$ are equivalent to (\ref{e:max1})--(\ref{e:j4}),
and yield exactly the same solution as above.
Writing the Lagrangian in terms of $\mu$ and $\bs{v}_{\text s}$ does
not yield any practical advantage, but clearly illustrates that gauge
invariance has become irrelevant---it is not broken, but has simply
left the stage.  Since all the fields are gauge invariant,
(\ref{e:l2}) does not even provide a framework to think of a spontaneous
violation of a gauge invariance.

These considerations apply to every field theory which displays the
Higgs mechanism.  In any such theory, the Lagrange density is invariant
under a global physical symmetry for a matter field, and invariant
under a local gauge symmetry, which affects both the matter field and
the gauge field.  The global symmetry is ``physical'' as we can
classify the states of matter according to their transformation
properties, while the gauge symmetry is ``unphysical'' as gauge
transformations have no effect on the states of matter, but only on
our description of these states.  In our example of a superfluid,
charged or neutral, the global symmetry transformation is 
\begin{equation}
  \label{e:global}
  \phi(x)\rightarrow\phi(x)+\lambda,
\end{equation}
where $\lambda$ is independent of spacetime.  This symmetry is
spontaneously violated, which means that there are many degenerate
ground states which map into each other under (\ref{e:global}).  For a
neutral superfluid, we obtain a massless mode according to Goldstone's
theorem.  The situation is more subtle for a superconductor, as the
matter field is coupled to a gauge field and the Lagrange density is
also invariant under the gauge transformation (\ref{e:gauge}).  This
``unphysical'' symmetry, however, seems to contain the physical
symmetry as the special case
\begin{equation}
  \label{e:globalgauge}
  \Lambda(x)=-\frac{\hbar c}{e^*}\lambda.  
\end{equation}
{The formal equivalence of the transformation (\ref{e:global}) and
(\ref{e:gauge}) with (\ref{e:globalgauge}) is 
at the root of the widely established but incorrect interpretation of
(\ref{e:global}) as a gauge transformation, and in particular of the
spontaneous violation of (\ref{e:global}) as a spontaneous violation
of a gauge symmetry.}  (This is presumably the reason why particle
physicists like Steven Weinberg speak of ``spontaneously broken gauge
symmetries'' interchangeably with ``the Higgs mechanism''.)  The
problem here is that the equivalence is only formal.  The gauge
transformation (\ref{e:globalgauge}) represents a transformation of
our description, similar to a rotation of a coordinate system we use
to describe a physical state, while the transformation
(\ref{e:global}) corresponds to a transformation of our physical
state, like a rotation of a physical system.  Clearly, a
(counterclockwise) rotation of the coordinate system has the same
effect on our equations as a (clockwise) rotation of the physical
system we describe with these equations, but the transformations are
all but equivalent.  It is hence incorrect to refer to the spontaneous
violation of (\ref{e:global}) as a spontaneous violation of gauge
symmetry.  A gauge symmetry cannot be spontaneously violated as a
matter of principle.  

The difference between the ``physical'' symmetry (\ref{e:global}) and
the gauge symmetry (\ref{e:gauge}) can also be appreciated at the
level of conservation laws.  The former yields particle number (or
charge) as a conserved quantity, according to (\ref{e:cont}), while
there is no conservation law associated with the latter.  In the
literature, (\ref{e:global}) is often referred to as a global gauge
transformation, and the conservation of charge attributed to gauge
invariance.  This view, however, is not consistent.  If one speaks of
a global gauge symmetry, this symmetry has to be a proper subgroup of
the local gauge symmetry group.  The alleged global gauge symmetry
hence cannot be a ``physical'' symmetry while the local gauge symmetry
is an invariance of description, or be spontaneously violated while
the local symmetry is fully intact.  The difference between the global
phase rotation (\ref{e:global}) and a global gauge rotation
(\ref{e:globalgauge}) is even more at evident at the level of quantum
states.  The BCS ground state (\ref{e:bcs}) is, for example, not
invariant under (\ref{e:global}), while it is fully gauge invariant,
as we have seen in Section \ref{sec:gau}

The conclusions regarding the physical significance (or maybe better
insignificance) of gauge transformations we reached here for
superconductors hold for any field theory which displays the Higgs
mechanism.

\section{QUANTUM EFFECTS}
%\section{Quantum effects}
\label{sec:qua}

This discussion of the Higgs mechanism applies only to simply
connected superconductors.  If we have a nontrivial topology or more
than one superfluid, the phase field $\phi$ reassumes physical
significance through its compactness, that is, the fact that its value
is only defined modulo $2\pi$.  In these situations, we are not
allowed to set $\phi(x)=0$ in the equations of motion or eliminate it
from the effective Lagrange density via (\ref{e:newfields}) or
(\ref{eq:a'}), as we would loose the information regarding the
compactness.  Since the phase field $\phi$ is multiplied by Planck's
constant $\hbar$ whenever it enters in the Lagrange density, any
effect due to the compactness of $\phi$ will depend on $\hbar$, and
only exist for $\hbar\ne 0$.  Therefore we refer to them as ``quantum
effects''.

The simplest of these effects in superconductors is the quantization
of magnetic flux, which is analogous to the quantization of vorticity
in neutral superfluids.  The effect was predicted by London in a
footnote in his first book~\cite{london,london2} almost a decade
before BCS proposed their microscopic theory.  Consider a macroscopic
superconductor with a hole in it, which may either be a hole in the
superconducting material or a line defect or vortex in the
superconducting order parameter.  Like in the case of a vortex in a neutral
superfluid, the phase field $\phi$ has to be single valued everywhere 
in the superconductor, but may change by a multiple of $2\pi$ as we
circumvent the hole or defect along a closed curve $\partial S$:
\begin{equation}
  \label{e:vortex2}
  \oint_{\partial S}\!\nabla\phi(x)d\bs{l}=2\pi n,
\end{equation}
where $n$ is an integer.  We now take $\partial S$ well inside the
superconductor, that is, separated at each point by a distance much
larger than the penetration depth $\lambda_{\text L}$ from the hole or
defect.  Then, according to the Meissner effect or our derivation of
London's equation above, which still applies locally, the superfluid
velocity
 \begin{displaymath}
  \bs{v}_{\text s}
  =\frac{1}{m^*}\Bigl(\hbar\nabla\phi+\frac{e^*}{c}\bs{A}\Bigr)
\end{displaymath}
has to vanish along $\partial S$, and (\ref{e:vortex2}) implies 
\begin{equation}
  \label{e:vortex3}
  \oint_{\partial S}\!\bs{A}(x)d\bs{l}
  =\int_{S}\!\bs{B}(x)\!\cdot\!\bs{n} da
%  =\frac{2\pi\hbar c}{e^*} n,
  =\frac{h c}{e^*}\cdot n,
\end{equation}
where we have used Stokes theorem once more.  The magnetic flux through the
hole or vortex is hence quantized in units of $hc/e^*$, which for
$e^*=2e$ is half of the Dirac flux quantum.  Note that the vorticity 
(\ref{e:vorti}) is not quantized in a superconductor.

\begin{figure} %[tb]
  \begin{center}
    \vspace{2mm}
    \psfrag{vortex}{vortex}
    \psfrag{(a)}{(a)}
    \psfrag{(b)}{(b)}
    \includegraphics[width=\columnwidth]{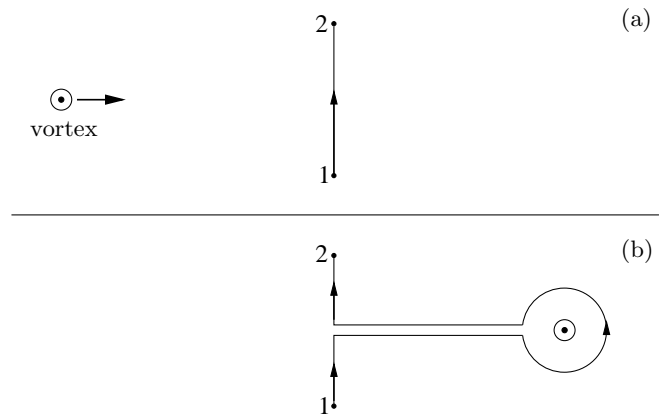}
  \end{center}
  \caption{Phase slippage: A vortex moving in a superfluid induces a 
    transverse gradient in the (electro) chemical potential by
    dragging a branch cut in the phase of the order parameter with
    it.}
  \label{f:slip}
\end{figure}
We now review two further quantum effects, which are similar in
neutral and charged superfluids; as in our derivation of the effective
theory above, the equations for the latter case contain the former as
the special case $e^*=0$.  One of the effects is phase
slippage~\cite{phil66}.  Consider two points $1$ and $2$ in a
superfluid, which are connected by a vertical path (see
\mbox{Fig.~\ref{f:slip}a).}  Now imagine we adiabatically move a vortex
from the very far left across the path to the very far right.  This
process yields a difference in the electrochemical potential between
the two points, which is according to (\ref{e:muelch}) given by
\begin{equation}
  \label{e:deltamu}
  \Delta\mu_{\scriptscriptstyle\text{el.chem.}}=
  -\partial_t \int_{1}^{2}\! 
  \Bigl(\hbar\nabla\phi(x)+\frac{e^*}{c}\bs{A}(x)\Bigr) d\bs{l}.
\end{equation}
where the line integral is taken along the path between the points.
The time integrated difference in the electrochemical potential is
hence given by the difference between the line integral at the end of
the process and the line integral at the beginning. %,
Let us first consider the case of a neutral superfluid, \ie $e^*=0$.
The line integral of $\nabla\phi$ will have changed by $2\pi$, as the
difference in the paths is topologically equivalent to encircling the
vortex once (see \mbox{Fig.~\ref{f:slip}b)}. 
Alternatively, we may say the vortex has dragged a branch cut
across the path.  (We assume that at the beginning and the end,
the vortex is so far away from points $1$ and $2$ that 
we can neglect its influence on the line integral.)
If we now have a continuous flow of vortices across the path, 
the line integral will pick up a contribution of $2\pi\hbar$
from each of them, and we obtain a chemical potential difference
\begin{equation}
  \label{e:phslip}
  \Delta\mu = h\langle\partial_t 
  N_{\text{v}}\rangle_{\scriptscriptstyle\text{av}}, 
\end{equation}
where $\langle\partial_t N_{\text{v}}\rangle_{\scriptscriptstyle\text{av}}$ is
the average rate of vortices crossing the path.

Let us now turn to the case of a superconductor, where we assume that
during the entire process the distance between the vortex and either
of the points $1$ and $2$ is much larger than the penetration
depth.  Since the line integral we obtain when encircling a
superconducting vortex along a circle well inside the superconductor
is zero,
\begin{equation}
  \label{e:encircle}
  \oint_{R\gg\lambda_{\text L}}\!\! \, 
  \Bigl(\hbar\nabla\phi(x)+\frac{e^*}{c}\bs{A}(x)\Bigr)d\bs{l}=0,
\end{equation}
we do not obtain a difference in the electrochemical potential as we
move an isolated vortex carrying a magnetic flux quantum across the
path.  So, at first sight, it may appear as there is no phase slippage
effect in superconductors.  The situation just described, however, is
not the general one, as we dragged a unit of magnetic flux with the
vortex from the very far left to the very far right.  This produced a
Hall effect which exactly canceled the phase slippage effect.  If we
consider a situation where we have a large, almost uniform magnetic
field and an Abrikosov vortex lattice or liquid in which the distance
between the vortices is much smaller than the penetration depth, and
we have a flow of vortices across the path, the magnetic field will
remain to a reasonable approximation unaffected by the flow and we
recover (\ref{e:phslip}) for the electrochemical potential difference.
The voltage we measure between the two points is then given by
\begin{equation}
  \label{e:nenrst}
  U=\frac{h}{2e}\langle\partial_t 
  N_{\text{v}}\rangle_{\scriptscriptstyle\text{av}}.
\end{equation}
This voltage is known as the Nernst effect in superconductors.

The last and possibly most striking quantum effect we review is the
Josephson effect~\cite{jos}.
Consider two superfluids or superconductors S$_1$ and S$_2$, which are
weakly coupled, say by a narrow constriction for superfluid particles
or a tunneling barrier for Cooper pairs. % (see Fig.\ \ref{fig:jos}).
The only requirement for the effect is that there is an energy
associated with the weak link, which depends on the (gauge invariant)
phase difference $\Delta\phi$ between two points 2 and 1 in
superfluids S$_2$ and S$_1$:
\begin{equation}
  \label{e:josen}
  E_{\text{junction}}=f(\Delta\phi) 
\end{equation}
with 
\begin{equation}
  \label{e:josph}
  \Delta\phi \equiv
  \phi(2)-\phi(1)+\frac{e^*}{\hbar c}\int_{1}^{2}\! \bs{A}(x) d\bs{l},
\end{equation}
where the line integral is taken along the path the superfluid
particles take when they move from one superfluid to the other.  Note
that $\Delta\phi$ divided by the distance between the points 2 and 1
is just the discrete version of the gradient term
\begin{displaymath}
  \hbar\nabla\phi(x)+\frac{e^*}{c}\bs{A}(x)
\end{displaymath}
we already encountered in the Ginzburg--Landau free energy, where the
magnetic energy (\ref{e:fmag}) was essentially given by its square.
We assume that $E_{\text{junction}}$ is likewise minimal for
$\Delta\phi=0$, which implies that the first term in a Taylor
expansion around this minimum is quadratic in $\Delta\phi$.  In the
case of the junction, however, this term is not sufficient.  Since
$\phi$ is only defined modulo $2\pi$, $f(\Delta\phi)$ has to be a
periodic function of $\Delta\phi$.  Josephson has shown that to a
reasonable approximation, it is given by
\begin{equation}
  \label{e:josf}
  f(\Delta\phi)=-E_0 \cos(\Delta\phi).
\end{equation}
Let us now assume a situation where both macroscopic superfluids are
in a state of equilibrium, but the phases are not necessarily aligned
relative to each other.  Then only the energy stored in the junction
depends on the phases of the superfluids, and the ``characteristic
equation'' (\ref{e:h2''}) becomes for superfluid S$_2$
\begin{equation}
  \label{e:josh2}
  \hbar\:\!\partial_t N_2
  =\frac{\partial E_{\text{junction}}(\Delta\phi)}{\partial\phi(2)}\:\!,
\end{equation}
where $N_2$ is the number of superfluid particles or Cooper pairs in
S$_2$.  (We would also obtain a similar equation for S$_1$, but since we
assume $N_1+N_2=\text{const.}$ and $E_{\text{junction}}$ only depends
on $\phi(2)-\phi(1)$, it does not provide any additional information.)  
The particle current from superfluid S$_1$ to S$_2$ is hence given by 
\begin{equation}
  \label{e:josj}
  J_{1\rightarrow 2}=\frac{1}{\hbar}
  \frac{\partial E_{\text{junction}}(\Delta\phi)}{\partial\phi(2)}
  =\frac{1}{\hbar}E_0 \sin(\Delta\phi).
\end{equation}
On the other hand, since the other ``characteristic
equation'' (\ref{e:h1'}) holds for each superfluid,
\begin{equation}
  \label{e:h1''}
  \hbar \partial_t \bigl(\phi(2)-\phi(1)\bigr)
  =-\bigl(\mu(2)-\mu(1)\bigr)+e^* \bigl(\Phi(2)-\Phi(1)\bigr).
\end{equation}
If we add  
\begin{displaymath}
  \partial_t \frac{e^*}{c}\int_{1}^{2}\! \bs{A}(x)d\bs{l}
\end{displaymath}
to both sides of (\ref{e:h1''}), we obtain
\begin{equation}
  \label{e:josdt}
  \hbar \partial_t \Delta\phi
  =-\bigl(\mu(2)-\mu(1)\bigr)-e^*\int_{1}^{2}\! \bs{E}(x)d\bs{l}
  =-\Delta\mu_{\scriptscriptstyle\text{el.chem.}}, 
\end{equation}
% On the other hand, if we integrate the other ``characteristic
% equation'' (\ref{e:muelch}) across the junction, we obtain
% \begin{equation}
%   \label{e:josdt}
%   \hbar \partial_t \Delta\phi
%   =-\Delta\mu_{\scriptscriptstyle\text{el.chem.}}, 
% \end{equation}
or, if we take $\Delta\mu_{\scriptscriptstyle\text{el.chem.}}$ time
independent,
\begin{displaymath}
  \Delta\phi(t)= 
  -\frac{\Delta\mu_{\scriptscriptstyle\text{el.chem.}}}{\hbar}\cdot t
%   + \text{const.}
%   +\Delta\phi(t=0)
   +\Delta\phi_0
\end{displaymath}
Substitution into (\ref{e:josj}) yields
\begin{equation}
  \label{e:josj'}
  J_{1\rightarrow 2}=\frac{E_0}{\hbar}
%  \sin(2\pi\nu t+\text{const.}),
%  \sin\bigl(2\pi\nu t+\Delta\phi(t=0)\bigr),
  \sin\bigl(2\pi\nu t+\Delta\phi_0\bigr),
\end{equation}
where 
\begin{equation}
  \label{e:josfreq}
  \nu\equiv-\frac{\Delta\mu_{\scriptscriptstyle\text{el.chem.}}}{h}
\end{equation}
is the Josephson frequency.  This implies that if the electrochemical
potential is equal for both superfluids, we find a DC particle current
depending on the initial alignment of the phases.  If there is a
difference in the potential, however, the current will oscillate with
frequency $\nu$.  This is called the AC Josephson effect.  The effect
exists for both neutral and charged superfluids, but it is much easier
to measure in a superconductor, as we can realize a difference in the
electrochemical potential by applying a voltage $U$ across the
junction, $\Delta\mu_{\scriptscriptstyle\text{el.chem.}}=-2eU$, and
easily measure oscillations in the electrical current.

Note that the Josephson effect, so astonishing its phenomenology may
be, follows through the ``characteristic'' equations of superfluidity
directly from the fact that there is a broken symmetry in
superfluids and that the compact phase field which labels the
different degenerate ground states is the field conjugate to the
density in the superfluid.  
% The other assumptions we made in this article, that the order
% parameter carries charge in the case of a superconductor, that there
% exists a linearly dispersing sound mode in neutral superfluids, which
% disappears due to the Higgs mechanism in charged superfluids, and that
% both current and momentum are carried by the same species of
% particles, was not required to explain any of the quantum effects.
The other assumption we made in this article, the assumption that both
current and momentum are carried by the same species of particles in a
superfluid, was not required to explain any of the quantum effects.

%\newpage
\section*{APPENDIX}

In this appendix, we 
%We now 
%
%explicitly show that in a BCS state all the electron pairs have
%condensed into one and the same two-particle state.
%
derive the position space wave function (\ref{e:bcsn}) by
projecting the BCS state (\ref{e:bcs}) onto a fixed number of pairs $N$.
%We omit factors of normalization. The BCS state may be written 
%
The (unnormalized) BCS state may be written
\begin{eqnarray}
  \nonumber \ket{\psi_\phi} \!&\!=\!&\!
  \prod_{\bs{k}}\left(1 +  e^{i\phi} \frac{v_{\bs{k}}}{u_{\bs{k}}} 
    c_{\bs{k}\up}^\dagger\,c_{-\bs{k}\dw}^\dagger\right)\vac
%  \\ \nonumber\rule{0pt}{18pt}  \!&\!=\!&\!
\end{eqnarray}
\pagebreak
\begin{eqnarray}
  \nonumber \phantom{\ket{\psi_\phi}} \!&\!=\!&\!
  \prod_{\bs{k}}\exp\Bigl(e^{i\phi}\frac{v_{\bs{k}}}{u_{\bs{k}}} 
    c_{\bs{k}\up}^\dagger\,c_{-\bs{k}\dw}^\dagger\Bigr)\vac 
  \\ \nonumber\rule{0pt}{18pt}  \!&\!=\!&\!
  \exp\Bigl(e^{i\phi} \sum_{\bs{k}} \frac{v_{\bs{k}}}{u_{\bs{k}}} 
    c_{\bs{k}\up}^\dagger\,c_{-\bs{k}\dw}^\dagger\Bigr)\vac
  \\ \nonumber\rule{0pt}{18pt}  \!&\!=\!&\!
  \exp\bigl(e^{i\phi} b^\dagger\bigr)\vac \!.
\end{eqnarray}
The pair creation operator $b^\dagger$ is given by
\begin{eqnarray}
  \nonumber
  b^\dagger\!&\equiv&\!\sum_{\bs{k}} \frac{v_{\bs{k}}}{u_{\bs{k}}} 
  c_{\bs{k}\up}^\dagger\,c_{-\bs{k}\dw}^\dagger
  \\ \nonumber\rule{0pt}{18pt}  \!&=&\!
  \int\! d^{3\,}\!\bs{x}_1 d^{3\,}\!\bs{x}_{2}\,\varphi(\bs{x}_1-\bs{x}_2)\,
  \psi^\dagger_\up(\bs{x}_1)\psi^\dagger_\dw(\bs{x}_2)\vac \!,
\end{eqnarray}
where $\varphi(\bs{x})$ is given by (\ref{e:varphi}).  If we now project
out a state with $N$ pairs according to (\ref{e:project}), we obtain
\begin{displaymath}
  \ket{\psi_N}=
%  \frac{1}{2\pi}
  \int_0^{2\pi}\!\!d\phi\,e^{-iN\phi}\exp\bigl(e^{i\phi}b^\dagger\bigr)\vac=
  \frac{2\pi}{N!}\;\!
  \bigl(b^\dagger\bigr)^N \vac \!,
\end{displaymath}
% \begin{eqnarray}
%   \nonumber
%   \ket{\psi_N} 
%   \!&\!\!=\!\!&\! 
%   \int_0^{2\pi}\!\!d\phi\,e^{-iN\phi}\exp\bigl(e^{i\phi}b^\dagger\bigr)\vac 
%   \\ \nonumber\rule{0pt}{18pt}  \!&\!\!=\!\!&\!
% %  \frac{1}{N!} 
%   \bigl(b^\dagger\bigr)^N \vac
% \end{eqnarray}
which is (up to a normalization) equivalent to (\ref{e:bcsn}).

\section*{ACKNOWLEDGMENTS}

I wish to thank T.\ Kopp, 
D.\ Schuricht, M.\ Vojta, and P.\ W\"olfle for discussions of various 
aspects of this work.
%fundamental aspects of superconductivity.

%\newpage
%\bibliographystyle{apsrmp}
%\bibliography{../bib/book,../bib/paper,../bib/schuricht,../bib/martin}

\begin{thebibliography}{99}
\bibitem{simp} For simplicity, we consider only time independent gauge
  fields and transformations.
\bibitem{conn} Note, however, that there is no connection as of yet 
  between the effective mass $m^*$ in (\ref{e:tvs})--(\ref{e:gal2})
  and $m^*$ in the effective Lagrange density.  The equivalence is
  only established with (\ref{e:js}) and (\ref{e:gww}) below.
\bibitem{phila} Anderson~\cite{phil66} refers to our definition
  (\ref{e:vs}) of $\bs{v}_{\text s}(x)$ as a ``pseudo-identity'', as
  the framework of his analysis does not provide the means to
  recognize the approximative nature of (\ref{e:j}).
\bibitem{london1} Strictly speaking, this is London's second and more
  famous equation.  The first equation,
  \begin{displaymath}
%    \partial_t \bs{J}_{\text{el.}}=\frac{{e^*}^2 n_{\text s}}{m^*}\bs{E},
    \partial_t \bs{J}=-\frac{e^* n_{\text s}}{m^*}\bs{E},
  \end{displaymath}
  is obtained from (\ref{e:muelch}) with (\ref{e:j3}) if we neglect the
  $\nabla\mu$ term, which is a rather crude approximation.
\bibitem{london2} See page 152.  Note that this prediction shows that
  London had fully understood the significance of vector potentials in
  quantum mechanics almost a decade before the acclaimed work by
  Y.\ Aharonov and D.\ Bohm.
\end{thebibliography}
%\end{document}

\end{document}